\documentclass{jfm}

\usepackage[latin1]{inputenc}
\usepackage[english]{babel}
\usepackage{float}
\usepackage{subfig}
\usepackage{textcomp}
\usepackage{amsmath}
\usepackage{graphicx}
\usepackage{setspace}
\usepackage{esint}
\usepackage{natbib}
\usepackage{color,soul}

\usepackage{psfrag}

\usepackage[T1]{fontenc}               
\usepackage[section]{placeins}         

\usepackage{fancyhdr}
\usepackage{mathrsfs}
\usepackage{amsfonts}
\usepackage{lscape}
\usepackage{longtable}
\usepackage{nicefrac}

\usepackage{array}
\newcolumntype{L}[1]{>{\raggedright\let\newline\\\arraybackslash\hspace{0pt}}m{#1}}
\newcolumntype{C}[1]{>{\centering\let\newline\\\arraybackslash\hspace{0pt}}m{#1}}
\newcolumntype{R}[1]{>{\raggedleft\let\newline\\\arraybackslash\hspace{0pt}}m{#1}}

\newcommand{\lyxdot}{.}

 \ifCUPmtlplainloaded \else
   \checkfont{eurm10}
   \iffontfound
     \IfFileExists{upmath.sty}
       {\typeout{^^JFound AMS Euler Roman fonts on the system,
                    using the 'upmath' package.^^J}%
        \usepackage{upmath}}
       {\typeout{^^JFound AMS Euler Roman fonts on the system, but you
                    dont seem to have the}%
        \typeout{'upmath' package installed. JFM.cls can take advantage
                  of these fonts,^^Jif you use 'upmath' package.^^J}%
       }
   \else
   \fi
 \fi

 \ifCUPmtlplainloaded \else
   \checkfont{msam10}
   \iffontfound
     \IfFileExists{amssymb.sty}
       {\typeout{^^JFound AMS Symbol fonts on the system, using the
                 'amssymb' package.^^J}%
        \usepackage{amssymb}%
          \let\leq=\leqslant
          
       }{}
   \fi
 \fi

 \ifCUPmtlplainloaded \else
   \IfFileExists{amsbsy.sty}
     {\typeout{^^JFound the 'amsbsy' package on the system, using it.^^J}%
      \usepackage{amsbsy}}
     {}
 \fi

\newsavebox{\astrutbox}
\sbox{\astrutbox}{\rule[-5pt]{0pt}{20pt}}

\newcommand{\dt}[1]{ \dfrac{  \partial{#1} }{\partial{t}   } }

\title[On a unified breaking onset threshold for gravity waves]{On a unified breaking onset threshold for gravity waves in deep and intermediate depth water}                                                      

\author[X. Barthelemy, M.L. Banner, {W.L. Peirson}, {F. Fedele}, {M. Allis}, {F. Dias}]{X. Barthelemy$^{1,2}$ \thanks{Email address for correspondence: x.barthelemy@unsw.edu.au}, %
 {M.L. Banner$^1$},  {W.L. Peirson$^2$},  {F. Fedele$^3$}, {M. Allis$^2$}  and {F. Dias$^4$} }

\affiliation{$^1$School of Mathematics and Statistics, UNSW Australia, 
$^2$Water Research Laboratory, School of Civil and Environmental Engineering, UNSW Australia, Sydney NSW 2052, Australia\\
$^3$School of Civil and Environmental Engineering, Georgia Institute of Technology, Atlanta, GA 30332, USA.\\
$^4$School of Mathematics and Statistics, University College Dublin, Ireland}

 \pubyear{2010}
 \volume{650}
 \pagerange{119--126}
 \date{?; revised ?; accepted ?. - To be entered by editorial office}
\begin{document}


\maketitle
\begin{abstract}
We revisit the classical but as yet unresolved problem of predicting the breaking onset of 2D and 3D irrotational gravity water waves.
This study focuses on domains with flat bottom topography and conditions ranging from deep to intermediate depth (depth to wavelength ratio from $1$ to $0.2$). Our calculations based on a fully nonlinear boundary element model investigated geometric, kinematic and energetic differences between maximally recurrent and marginally breaking waves in focusing wave groups. Maximally steep non-breaking (maximally recurrent) waves are clearly separated from marginally breaking waves by their normalised energy fluxes localized near the crest region. On the surface, this reduces to the local ratio of the energy flux velocity (here the fluid velocity) to the crest point velocity for the tallest wave in the evolving group. This provides a robust threshold parameter for breaking onset for 2D and 3D wave packets propagating in uniform water depths from deep to intermediate. 
Warning of imminent breaking onset was found to be detected up to a fifth of a carrier wave period prior to a breaking event. 
\end{abstract}

 \begin{keywords} Authors should not enter keywords on the manuscript,
 as these must be chosen by the author during the online submission
 process and will then be added during the typesetting process (see
 http://journals.cambridge.org/data/\linebreak[3]relatedlink/jfm-\linebreak[3]keywords.pdf
 for the full list) \end{keywords}

\section{Introduction\label{sec:Introduction}}

Despite its long research history, the physics underpinning the breaking of water waves has remained incompletely understood, including prediction of its onset and strength. Yet this knowledge is of fundamental importance in quantifying atmosphere-ocean exchanges, determining structural loadings on ships and platforms, and optimising operational strategies for maritime enterprises.

Many criteria for predicting breaking onset of 
water waves have been proposed since the pioneering study of \citet{Stokes1847}.
These criteria arise from theoretical arguments based on idealized models, numerical simulations, laboratory experiments and field observations.  However, while adding many insights, these approaches have not yielded a robust breaking threshold for phase-resolved waves in the physical domain, reflecting the complexity  of the underlying dynamical processes. In fact, there is a glaring absence of a precise definition of breaking onset.

Briefly, it has long been considered that breaking is a process with a threshold, with criteria for predicting breaking onset falling into three categories: geometric, kinematic and energetic. The majority of breaking criteria have been based on a geometric or kinematic threshold, and mainly limited to plane (2D) waves. Geometric threshold variables have included wave steepness, wave asymmetry, maximum theoretical (global) steepness and the occurrence of a transient vertical segment on the forward face of the wave crest; kinematic threshold variables have included the Lagrangian crest acceleration and the ratio between crest fluid speed and phase speed. The
recent comprehensive review by \citet{Perlin2013} provides an excellent overview of the collective observational and
theoretical effort and outcomes based on kinematic/geometric approaches. The recent contributions of \cite{Shemer2013}, \cite{Shemer2014}, \cite{Kurnia2014} and \cite{Shemer2015} add to this otherwise exhaustive coverage. Overall, current knowledge does not support a kinematic or geometric criterion that provides a generic threshold which differentiates breaking from recurrent behaviour for deep water waves. 
While the vertical tangent segment and kinematic criteria provide valid \emph{a posteriori} conditions for breaking onset, they provide no dynamical insight or advance warning of imminent breaking. 


A third approach based on dynamical criteria has been explored to explain the onset of breaking. 
This concept is based on the 
evolution of the intra-group energy flux which causes the tallest crest of an unsteady wave group to break when a local stability threshold is exceeded. Monitoring the energy flux field in this highly nonlinear, unsteady flow environment makes rigorous analysis difficult. 
The overview article by \cite{Tulin2000} highlights the very insightful inroads made by Tulin and his collaborators over the previous decade into unsteady nonlinear wave group evolution and breaking, based on  intra-group energy flux theory, simulations, observations and analyses.  One of the key results they proposed from their studies is that  breaking onset is initiated within a wave group when a crest particle speed exceeds the linear group speed. Pending verification of its general validity, this criterion is able to signal breaking onset much earlier than the traditional kinematic criterion.  

Subsequently, \citet{Banner1998}, \citet{Song2002} and the experimental study of \citet{Banner2007}
investigated a growth rate based on a parametric energy convergence rate for 2D wave groups, using a frame of reference that tracks the
wave group maximum. \citet{Perlin2013} discuss the merits  of this approach based on the further study of \citet{Tian2008} for 2D wave breaking. Very recently, \citet{Derakhti2016} reported very encouraging support for this approach in their numerical study of unsteady 2D wave packets in a model framework that can accommodate sequential (multiple) breaking events as the packet evolves. They  confirmed the presence of systematic crest/trough leaning motions, as investigated in detail in \cite{Banner2014}. 
In the present study, we revisit this local energy growth rate approach for both 2D and 3D breaking onset simulations, for which our findings are given below in subsection 
\ref{sub:Dynamical-criteria}.


Finally, significant additional challenges arise in representing wave breaking in broad-banded directional sea states in the spectral (wavenumber-frequency) context. This is the domain used for computing ocean wave forecasts (e.g. see \citet{Chalikov2012}).  In that context, there is even less consensus on how to predict/identify breaking events. However, this is beyond the scope of the present paper, where we focus on wave breaking in the physical space-time domain. 



\section{Rationale underpinning our breaking onset threshold investigation}

Present understanding of the physics underpinning wave breaking onset in the physical domain is fragmentary, including a precise definition. This has precluded  reliable prediction of wave breaking onset even in controlled laboratory and numerical wave basin conditions. Our investigation directly addresses this time-honoured knowledge gap.

Based on energy flux considerations, we propose a new breaking onset threshold parameter. The behaviour of this parameter in the wave crest region is found to be of central importance. A new definition for a generic breaking onset threshold emerges naturally. Through an ensemble of simulations of diverse nonlinear wave packets in both deep and intermediate depth water over a flat bottom, we establish the existence of a breaking onset threshold band for this parameter, determining its upper and lower bounds. Below the lower bound of the proposed breaking threshold band, steepening carrier waves evolve through the packet envelope maximum without the occurrence of a vertical tangent in the wave  surface profile. All  maximally steep non-breaking carrier waves exist below this lower bound, which we also refer to as the maximum recurrence threshold.

  Thereafter, the smallest increment in the crest wave energy density (e.g., as produced by increasing the wave paddle amplitude) causes the breaking threshold parameter at the tallest crest to increase. After it evolves further, it exceeds our proposed breaking onset threshold and a significant change is initiated in the carrier wave crest appearance. Irreversible degeneration proceeds as illustrated in Figure 1 of \cite{Duncan2001}. As discussed in detail in sections 3.1 and 3.2 of \cite{Duncan2001}, the ensuing shape of an actively breaking crest depends on the wave scale, the influence of surface tension and on the strength of the breaking, which reflects the energy convergence rate at the wave crest. The strength of the breaking event is an unknown function of the magnitude of the breaking parameter above the threshold, with the smallest exceedance margins associated with very weak (i.e. marginal) breaking.

For long carrier wavelengths, the evolution to breaking leads to plunging jets of varying size relative to the wavelength:  the lower the energy input rate to the wave, the smaller the plunging
 jet. For short wavelengths, the jet formation is modified by surface tension and has been 
investigated in detail with boundary element calculations, theory, and experiments (e.g. \cite{Tulin2000}). When the wavelength exceeds about 2 m, breaking starts with the formation of a small plunging jet, just as it does when the surface tension vanishes. As the wavelength is decreased, surface tension forces become relatively larger and the jet tip becomes rounded. For wavelengths less than about 0.5 m, the jet is replaced by a bulge, and capillary waves appear upstream of the leading edge (toe) of the bulge, as shown in Figure 1 of Duncan (2001).  In the present investigation, surface tension is not included explicitly. However, we show below (in section \ref{sec:surf.ten}) that its estimated effect on our proposed generic breaking onset threshold is negligible for wavelengths longer than 1 m.

In this study, we used a fully nonlinear wave code capable of capturing the initial stage of crest overturning, including the critical visible signature of a transient vertical tangent on the forward face of the wave crest. The results from this code were used to determine the upper and lower bounds of our proposed breaking onset threshold band. Note that we do not directly solve the instability problem for breaking onset, nor does our model provide information beyond the initial stage of crest overturning for waves that exceed our breaking threshold. In common with many other studies, this choice of using strongly chirped wave packets which focus rapidly minimises the evolution time and fetch to breaking, or recurrence, and hence the CPU time. It also reduces the adverse Lagrangian drift implications for the computational grid (see end of section \ref{sec:NWT2} for details). However, the high chirp rate also restricted our attention to weak breaking cases, as further small increments in paddle amplitude only produced wave breaking at the paddle.


Our breaking onset parameter is formulated as the local energy flux relative to the local energy density, normalised by the local crest speed, and is operative throughout the subsurface region including the wave surface. For the condition of zero surface pressure, its projection on to the wave surface reduces to a simple quasi-kinematic form involving the ratio of the surface fluid speed $u$ to the crest point speed $c$. We note that the boundary geometry does not enter the breaking onset criterion explicitly, so it is potentially applicable to variable depth bottom topography scenarios. 

In the present paper, the behaviour of our breaking onset threshold parameter is determined from an ensemble of numerical simulations of 2D and 3D chirped focusing wave packets on deep and intermediate depth flat bottom topography, ranging down to one fifth of the dominant wavelength. For this class of wave packets, the results establish the existence of a generic narrow threshold band for breaking onset, for which $u/c$ is found to be appreciably lower than the traditional kinematic breaking criterion of $u/c > 1$. As foreshadowed above, the resemblance of the surface projection of our breaking onset threshold to a kinematic criterion is fortuitous. It only takes this form for zero surface pressure forcing conditions. Its intrinsic dynamical nature is confirmed by two additional factors:  the concomitant subsurface threshold does not reduce to a kinematic form and the breaking threshold $u/c$ ratio is considerably below unity. 
The findings from our numerical investigation for 2D deep and intermediate depth water waves are found to agree closely with measurements from  our companion observational studies by \citet{Saket2017} and \citet{Saket2017a}. Their observations also validated our proposed breaking onset threshold for moderate wind forcing and also for modulationally-focusing bimodal wave packets. 
\section{Methodology \label{sub:Discretisation} }

\subsection{Wave generation}

Wave groups, either 2D or 3D and in deep or intermediate depth water over flat bottom topography, were generated using a bottom-hinged flap-type snake wavemaker at one end of the tank. The motion $X_{p}(t,y)$ of the wavemaker flap at the lateral location $y$ followed the Class 3 'chirp packet' motion from \citet{Song2002} (this is implicit hereafter in the C3 designator in each documented run file):
\begin{multline}\label{chirpgroup}
X_p(t,y) =  -0.25 A_p \left(1+\tanh\left(\frac{4\omega_{p}t}{N\pi}\right)\right)\left(1-\tanh\left(\frac{4(\omega_{p}t-2N\pi)}{N\pi}\right)\right)\\
  \sin\left(\omega_{p}\left(t-\frac{\omega_{p}C_{ch}t^{2}} {2}\right)+\Phi\left(y,X_{conv},Y_{conv}\right)\right), 
\end{multline}
where $t$ is the time, $A_p$ is the amplitude of the paddle motion, $N$ is the number of waves in the temporal wave packet, $\omega_{p}$ is the baseline driving frequency of the paddle, with corresponding linear wavenumber $k_p$, 
and $C_{ch}=1.0112 \times  10^{-2}$ specifies the chirp rate used in this study. The phase $\Phi\left(y,X_{conv},Y_{conv}\right)$ specifies the coordinates of the point of linear convergence (see \citet{Dalrymple1988,Dalrymple1989}):
\begin{eqnarray}
\Phi(y,X_{conv},Y_{conv}) & = & k_p y \sin\theta(y) + k_p \left(X_{conv}\cos\theta(y)+Y_{conv}\sin\theta(y)\right)\\
\theta(y,X_{conv},Y_{conv}) & = & \arctan\left(\frac{y-Y_{conv}}{X_{conv}}\right),
\end{eqnarray}
where $\theta$ is the focal angle at location $y$ along the paddle. The downstream boundary opposite the wave paddle is a fully-absorbing boundary condition, as in \citet{Grilli1997}.

\subsection{Numerical wave tank} \label{sec:NWT2}

There has been growing interest in the development of three-dimensional models which inherently incorporate nonlinearity and associated dispersion effects. The broad--bandedness in both frequency and direction of real sea states poses significant challenges in numerical simulation. \citet{Bateman2001} demonstrated the importance of directionality and the consequent benefits of efficient wave modelling when comparing numerical simulations with the laboratory observations of \citet{Johannessen2001}. High--order spectral expansion approaches using efficient FFT solvers for application to 3D waves have been developed (e.g. \citet{Ducrozet2011}
and \cite{Fedele2016a}). Another option is to solve the full Navier--Stokes equations (\citet{Park2003}), but viscous flow solvers tend to be too dissipative and computationally time--consuming.

Numerical models of 3D potential flow wave propagation can be divided into three main categories: (a) boundary element integral methods (BEM): e.g. \citet{Baker1982}, \citet{Bateman2001}, \citet{Clamond2001}, \citet{Grilli2001}, \citet{Fochesato2007}, \citet{Guyenne2006}, \citet{XUE2001}, \citet{Hou2002}, \citet{Fructus2005}; (b) finite element method (FEM) e.g. \citet{Ma2001}; (c) spectral methods: e.g. \citet{Dommermuth1987}, \citet{West1987}, \citet{Craig1993}, \citet{Nicholls1998}, \citet{Bateman2001}.

Spectral methods based on perturbation expansions are known to be very efficient. These methods reduce the water-wave problem from one posed inside the entire fluid domain to one posed on the boundary alone, thus reducing the dimension of the formulation. This reduction can be accomplished by using integral equations over the boundary of the domain (so-called boundary integral methods) or by introducing boundary quantities which can be expanded as Taylor series for reference domain geometries (\citet{Xu2009}). Both approaches have been summarised recently by \citet{Ma2010}. BEM techniques are efficient for representing wave propagation and overturning until the wave surface reconnects (\citet{Grilli1996}).

The present study used a boundary element numerical wave tank (hereafter NWT) code called WSIM, which is a 3D extension of the 2D code developed by \citet{Grilli1989} to solve the single-phase wave motions of a perfect fluid. It has been applied extensively to the solution of finite amplitude wave propagation and wave breaking problems (see chapter 3 of \citet{Ma2010}).

The perfect fluid assumption makes WSIM unable to simulate breaking impact subsequent to surface reconnection.
However, its potential theory formulation enables it to simulate wave propagation in a CPU-efficient way, without the diffusion issues of viscous numerical codes. The simulation of wave generation and development of the onset of breaking events can be carried out with great precision, as shown by \citet{Fochesato2006}, \citet{Fochesato2007}.

WSIM has been validated extensively for wave evolution in deep and intermediate depth water and shows excellent energy conservation (\citet{Grilli1989}, \citet{Grilli1990}, \citet{Grilli1994}, \citet{Grilli1996}, \citet{Grilli1997}, \citet{Grilli2001}, \citet{Fochesato2004}, \citet{Fochesato2006}, \citet{Fochesato2007}). Its kinematical accuracy has been validated against the analytical solutions for infinitesimal sine waves in \citet{Phillips1977}.


WSIM uses a boundary element method (BEM) to compute field variables. The 16-node quadrilateral elements provide global third order precision, and high-order tangential derivatives needed for the time discretisation are computed in a local 25-node quadrilateral element curvilinear coordinate system giving fourth order precision. A fast multipole algorithm is used to invert the BEM problem.

\begin{figure}
\centering
\begin{tabular}{cc}
\subfloat[\label{fig:C3N5-waveprofile}\label{fig:C3N5-waveprofile}
]{\includegraphics[width=0.5\textwidth]{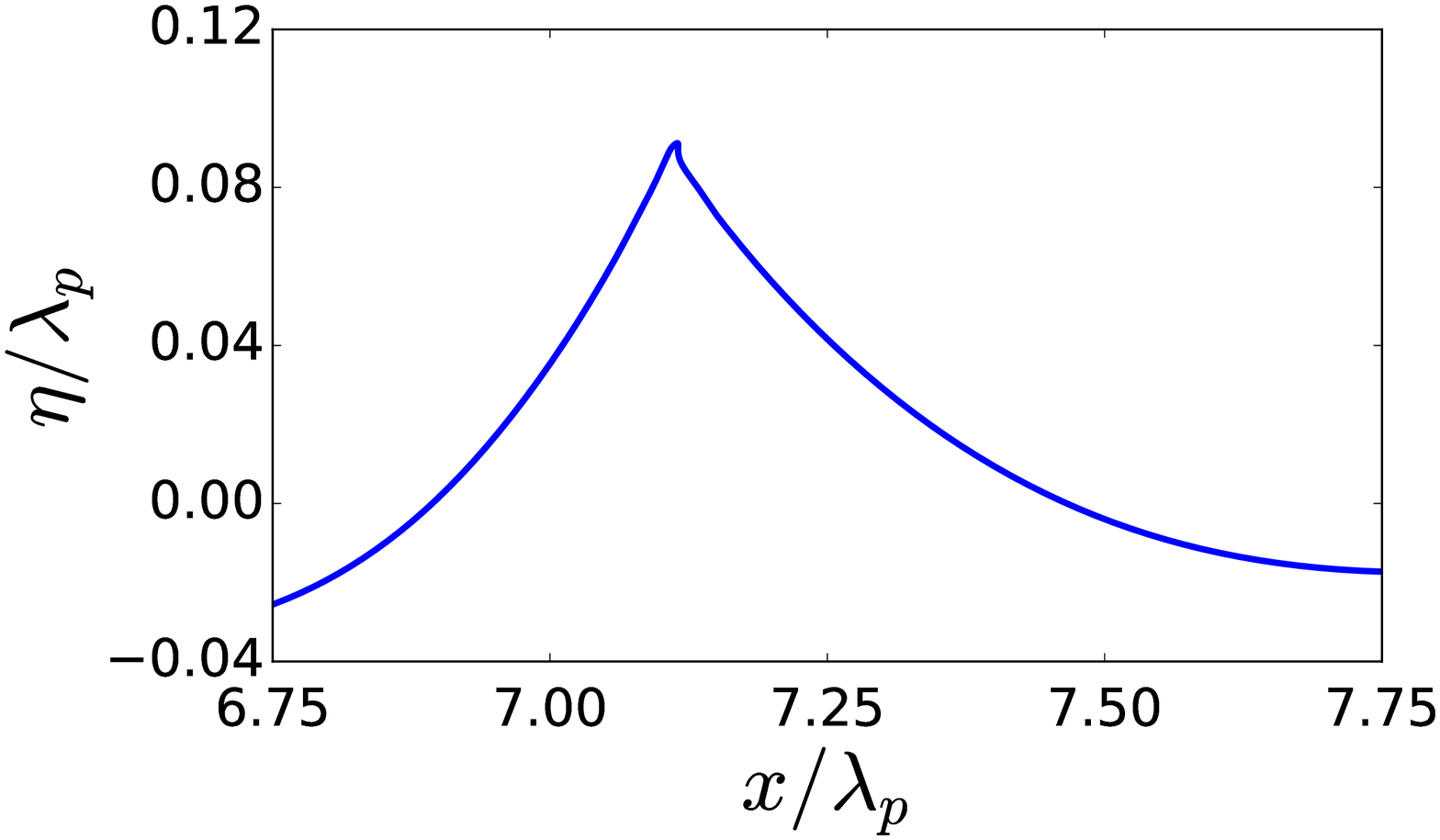}}  
&
{\subfloat[\label{fig:C3N5-waveprofile-Zoom} \label{fig:crest-sequence}
]{\includegraphics[width=0.5\textwidth]{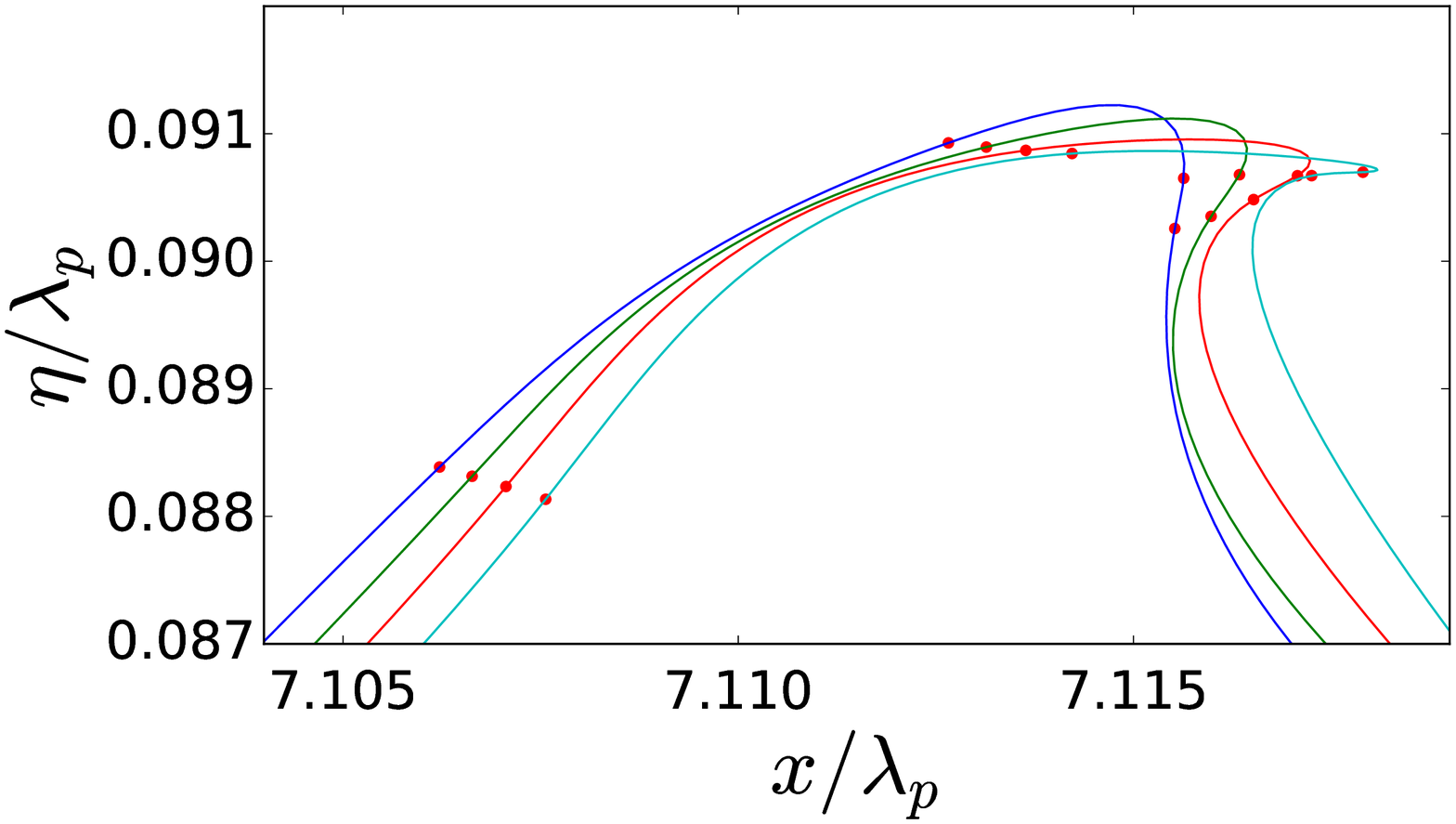} }}
\end{tabular}
\caption{Evolution details of 
the initiation of crest roll-over at the leading crest of a 2D C3N5 deep water wave packet undergoing weak breaking. The sequence of local crest surface profiles is shown in panel (b), starting after the appearance of the vertical tangent on the forward face. The crest tip moves from left to right. The small solid circles indicate the computational nodes. Details of the splined curves are given in section \ref{sec:NWT2}.
\label{fig:Wave-profiles}}
\end{figure}

3D simulations, with $x$ as the main direction of propagation, $y$ the transverse horizontal direction and $z$ the vertical direction, were run using mainly $16$ nodes per wavelength. The insensitivity of the results to the resolution was established by a subset of runs using $32$ nodes per wavelength. The number of nodes in the $y-$direction or in the $z-$direction was adjusted to keep the boundary element aspect ratio close to $1$. 
Figure \ref{fig:C3N5-waveprofile} shows a typical elevation profile of a gently breaking 2D C3N5 deep water wave packet just after exceedance of our breaking onset threshold. This case is discussed in detail in section \ref{sec:subsurface} and shown in figures \ref{fig:breakingthresohold2D-3Dview-B} and \ref{fig:breakingthresohold2D-2Dview-B}. Figure \ref{fig:crest-sequence} shows the sequence of elevation profiles of this overturning wave crest at incremental time steps. 
It highlights the ensuing development  of the crest tip jet following the occurrence of a vertical tangent on the forward face, confirming the overturning capability of WSIM.

\begin{figure}
\centering
\begin{tabular}{cc}
\subfloat[\label{fig:View-N5-A035-5}
]{\includegraphics[width=0.49\textwidth]{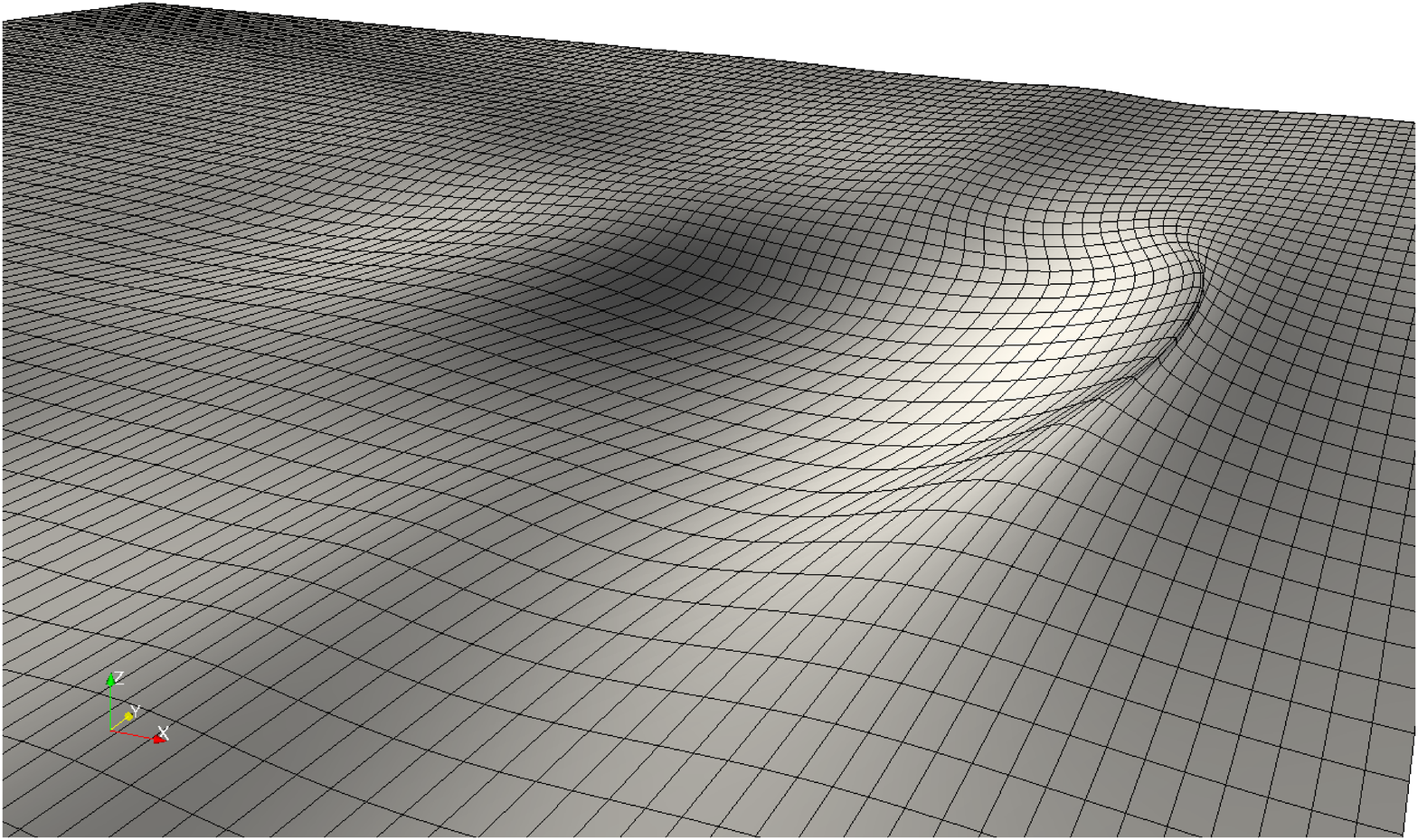}}  
&
{\subfloat[\label{fig:View-N5-A035-3}
]{\includegraphics[width=0.49\textwidth]{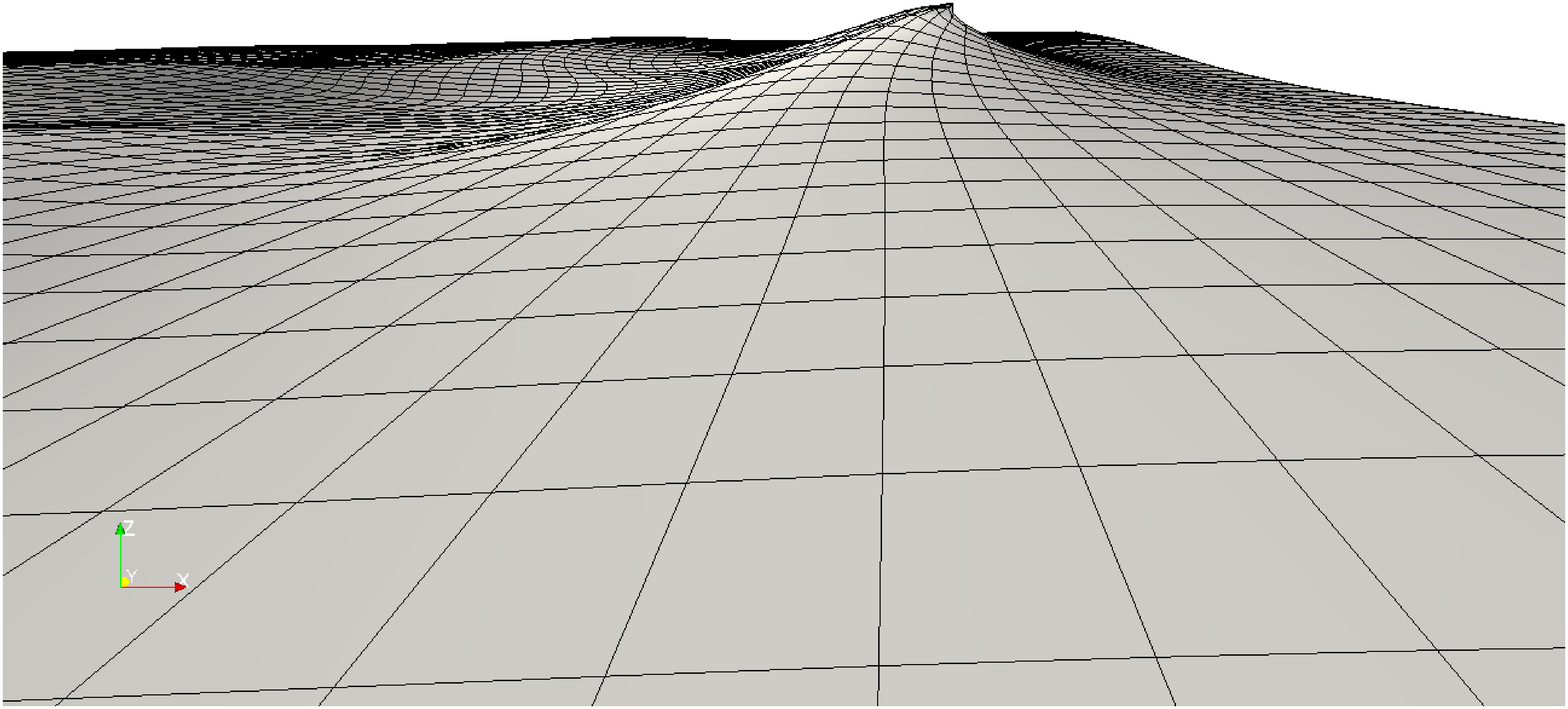}} }
\end{tabular}
\caption{Numerical wave tank simulation showing the initiation of breaking at the leading crest of a 3D converging C3N5 deep water wave packet as it propagates to the right. Panel (a) shows the computational domain while panel (b) shows a zoomed-in view of the overturning crest. The simulation used 16 nodes per wavelength, a paddle amplitude of $A_p / \lambda_p=0.2067$ and a linear focal distance of $x/ \lambda_p = 5$. The rectangular cells are artefacts of the visualisation. \label{fig:Breaking-zoom}}
\end{figure}

Figure \ref{fig:Breaking-zoom} shows a typical simulation using $16$ nodes per wavelength. It illustrates the breaking initiation of the crest of a 3D converging deep water chirped wave packet with $5$ waves in the temporal group. Even though each cell is represented by a flat quadrangle in the visualisations in figure \ref{fig:Breaking-zoom}, bi-cubic and fourth-order basis polynomials were used, respectively, for the physical variables and for the geometry, providing second-order curvature discretisation.  

Each breaking case we investigated conforms to this generic systematic progression (exceeding the breaking threshold, subsequent vertical tangent on forward face of crest and formation of initial crest tip jet). 
The numerics handles this smoothly, with the code stopping when the boundary elements at the crest tip jet become enmeshed. However, this occurs well beyond the exceedance of the breaking onset threshold and has no impact in determining this threshold. This is described in section \ref{sec:numcv} and in greater detail in appendix \ref{sec:detailCv}.  

The mixed Eulerian-Lagrangian numerical scheme and the high nonlinearity of the waves make the free surface mesh prone to distortion by the Lagrangian drift current. Extreme care was taken so that even at maximum recurrence, only a moderate Lagrangian drift was produced and the mesh did not deform significantly. An important benefit of the local Lagrangian drift is the clustering of the nodes around sharp crests. 


\subsection{Scaling parameters in the numerical simulations}

The numerical simulations use the following deep water (DW) non-dimensional parameter scalings: reference length scale is the wavelength $\lambda_{p}$ at the wavemaker (with corresponding wavenumber $k_p$); reference time scale is based on the baseline frequency $\omega_{p}$ of the paddle. The deep water dispersion relation imposes a gravitational acceleration $g_{DW}=2\pi$ and a reference linear phase speed $c_{DW}=1$.  Results have been produced for depth to wavelength ratios $(d/\lambda_p)$ belonging to the interval $[0.2,1]$.

\subsection{Numerical convergence} \label{sec:numcv}

This research aims to identify a criterion that can robustly predict whether growing 2D or 3D carrier waves in evolving wave groups will attain their maximum steepness without breaking (recurrent waves) or proceed to break, with overturning crests. To achieve this aim, the weakest form of breaking, marginal breaking, is computed and compared with the corresponding maximum recurrent case. 
We carried out a detailed sensitivity study 
that demonstrates convergence of the NWT model even for such low-intensity breaking.
Sensitivity tests were performed for one 2D ensemble, C3N5, using the two different resolutions described above (16 and 32 nodes per wavelength). 
We confirmed that both resolutions share the initial stage of breaking onset, confirmed by the occurrence of a vertical forward face segment and a multiple-valued free surface. 
While some minor differences were seen between the different resolutions, the numerical convergence of the computed breaking parameter is discussed in detail in appendix \ref{sec:detailCv}, where it is established that the resolution of 16 nodes per wavelength suffices to robustly quantify the breaking onset threshold.


\section{Analysis of simulation data using previous breaking criteria}

The recent review paper of \citet{Perlin2013} provides a detailed analysis of the different classes of proposed breaking criteria. Shortcomings have been identified in each of the criteria proposed to date. In the present study, several of these breaking criteria were also investigated using our simulation data and their validity evaluated. 

\subsection{Geometrical criteria }

\begin{figure}
\centering %
\includegraphics[width=0.8\textwidth]{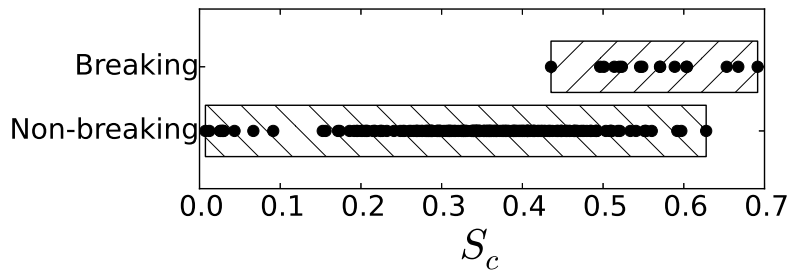} %
\caption{Distribution of local crest maximum steepness $S_c$ partitioned according to breaking or recurrent events, for all the C3 wave packets investigated in this study.  Note the extensive $S_c$ range shared by non-breaking and breaking waves.\label{fig:Steepness-Breaking}}
\end{figure}

As reported in \citet{Perlin2013}, geometric threshold criteria were not found to be robust (in the generic sense) in previous observational studies. Figure \ref{fig:Steepness-Breaking} provides an overview of the performance of the steepness criterion for breaking onset for the present data set. It shows the distribution of the maximum local crest steepness $S_{c}= \pi a / \lambda_{c}$ of each of the recurrent or breaking wave packets. Here $a$ is the crest amplitude and $\lambda_{c}$ is the horizontal extend, measured between the wave profile zero up and down crossings that span the crest (\citet{Banner2014}). Our results confirm previous findings that breaking onset cannot be discriminated by the steepness criterion since some breaking crest cases have a significantly lower $S_{c}$ than recurrent waves. 

\subsection{Kinematic criteria}\label{sec:kinematics}

As described in \citet{Perlin2013}, various kinematic breaking criteria have been proposed.
We note that \cite{Stansell2002} and \cite{Kurnia2014} mention a variety of purely kinematic criteria (ratio of fluid speed to wave speed larger than $0.7$, $0.8$ and $1$) which are embraced by our findings in the present paper.  However, there is no guidance as to the universality of the detection of breaking onset. 

Recently, \citet{Banner2014} investigated the influence of unsteadiness in dispersive wave packets, based on results from numerical simulations and complementary laboratory and ocean tower measurements. Their study highlighted the existence of a significant kinematic/geometric phenomenon attributable to the extra degrees of freedom in such unsteady wave packets. This results in an additional generic oscillatory crest/trough leaning mode characterising the carrier wave evolution. For focusing deep water wave packets, this manifests as a systematic crest speed slowdown of approximately $20\%$ of the linear phase velocity, reconciling why their breaking crests are observed to have initial speeds typically $20\%$ lower than the corresponding linear phase speed for that wavelength. The source of the crest slowdown mechanism is investigated in \citet{Banner2014}, \cite{Barthelemy2015} and \citet{Fedele2014}. Most significantly, it determines the underlying crest motion from which the breaking onset initiates, which is a key element in our new breaking framework.

\subsection{Dynamical criteria\label{sub:Dynamical-criteria}}

Dynamical criteria link the physics of breaking onset to the energy fluxes associated with the underlying unsteady wave group structure. Conceptually, the rate of convergence of intra-group energy flux exceeds a local stability level at a particular crest, which triggers this crest to break. The highly nonlinear, spatially non-uniform and unsteady nature of the flow field makes rigorous analysis difficult. 
One of the key results in section \ref{sec:Introduction} describing dynamical breaking criteria was the proposition that breaking onset is initiated when the crest particle speed exceeds the linear group speed (\citet{Tulin2000}).
We were able to investigate this criterion for chirped nonlinear wave packets for a range of depth conditions, including 3D cases. However, based on our findings in section \ref{sec:comparisonpreviouscriteria} below, our results do not support their proposed criterion.

We also revisited the wave group-related energy flux approach investigated 
in the modelling study of \cite{Song2002} and validated observationally in \cite{Banner2007}. This approach examined the possibility of a generic local parametric energy convergence rate for 2D wave groups following the wave group maximum (see section 3.3 in \citet{Perlin2013}). Further support was recently reported for 2D breaking onset by \citet{Derakhti2016} (section \ref{sec:Introduction}).   
However, two factors motivated our present search for an alternative breaking criterion based on an energy convergence rate threshold.

Firstly, our present study found that 2D and 3D breaking onset behaviour did not closely match the 2D threshold growth rate proposed by \citet{Song2002}. For both 2D and 3D cases in our study, the computed local carrier wavenumber used in constructing the \citet{Song2002} growth rate did not increase monotonically as the wave steepened, contrary to the analysis of \citet{Song2002}. This departure was also observed experimentally in section 6.3 of \citet{Allis2013}. As a result, the diagnostic growth rate trajectory departed significantly from the results reported in \citet{Song2002}. 
Secondly, the approach of \citet{Song2002} requires tracking, for any given crest, the 
space-time locus of its maximum elevation for at least 2 cycles prior to its reaching its ultimate maximum (either the recurrence maximum or breaking onset). Aside from its measurement complexity, this approach becomes tenuous for cases when more rapid approach to breaking onset occurs with fewer than 3 growth cycles. These factors underpinned our systematic 
search for a less restrictive energy flux-based  breaking criterion.

\section{New breaking criterion based on the local energy flux velocity}

Conceptually, the onset of breaking may be regarded as the inability of the waveform to accommodate a local wave energy flux which exceeds that in the corresponding maximum recurrent case. It is observed that breaking of the dominant waves typically occurs at the crest of the tallest dominant wave within a group, showing the preferred crest localisation of the phenomenon. This is consistent with the open ocean observations of \cite{Holthuijsen1986}. Excess local wave energy flux can arise from a variety of sources, such as intra-wave energy exchanges, wind-wave exchange, geometrical and temporal 3D wave focusing, wave-current interactions, among others. For the present focus on unforced water wave groups, we hypothesise that the same breaking onset physics could apply whether the wave group is evolving in deep water or in intermediate water depth. The shortcomings of the various criteria described above led us to focus on the role of excess wave energy flux as the underpinning element of a generic breaking criterion. 
The wave energy flux vector is defined in section (2.3) of \citet{Phillips1977}, together with a discussion of its local conservation equation (2.3.2).

\subsection{Energy flux considerations in nonlinear wave groups} \label{sec:energyfluxconsideration}

The mechanical energy balance equation relates the local rate of change of the energy density $E$
\[E=\rho g\left(z-z_0\right)+\frac{1}{2}\rho\Vert\mathbf{u}\Vert^{2}\]
to the divergence of the local energy flux $\mathbf{F}$
\[\mathbf{F}=\mathbf{u}\left(\left(p - p_0\right)+\rho g\left(z-z_0\right)+\frac{1}{2}\rho\Vert\mathbf{u}\Vert^{2}\right)\,\]
where $p$ is the pressure, $p_0$ is the ambient pressure above the surface (taken as zero without loss of generality), 
$\Vert\mathbf{u}\Vert$ is the fluid speed, $g$ is gravitational acceleration, $z$ is the vertical coordinate and $z_0$ the datum.


With the above definitions, equation (3.6.14) in \citet{Phillips1977} shows the conservation law for the depth-integrated energy. Based on (3.6.14),  we performed an analysis of the crest behaviour of the depth-integrated energy density, depth-integrated energy flux and its gradient at maximal focusing for representative nonlinear wave packets. 
We investigated examples of chirped packets with different numbers of carrier waves. Our aim was to determine whether breaking onset provides a distinctive signature within the depth-integrated energy context. We concluded that this depth-integrated approach obscures this apparently highly localised crest instability. To detect the transition to breaking above the background wave energy, it became evident that a local energy flux analysis in the neighbourhood of the crest was needed. This is described in the following section.

\subsection{Breaking criterion based on the local energy flux velocity\label{sub:Breaking-criterion-based}}

For the present purposes, in an inertial frame of reference, the local energy density conservation law (\citep{Phillips1977}, equation (2.3.2)) takes the Eulerian form:
\begin{align} 
      \dt{E} + \nabla \cdot \left(\mathbf{u}\left(E+p\right)\right)= &0  \label{eq:Phillips}
\end{align}                                                                	                                  
From equation (\ref{eq:Phillips}), in an inertial frame of reference, the energy flux velocity is seen to be the fluid velocity $\mathbf{u}$.  However, to gain a refined understanding of the energy flux to the tallest crest in an evolving nonlinear wave packet, the corresponding conservation equation for a control volume moving with the (unsteady) crest velocity $\mathbf{c}$ has the form (\citep{Tulin2007}, equation (1.2))
\begin{align}
\dfrac{D_cE}{Dt} + \nabla \cdot \left(\left(\mathbf{u}-\mathbf{c}\right)E+\mathbf{u}p\right)=& 0 \label{eq:Tulin}
\end{align}
where $\dfrac{D_cE}{Dt}= \dt{E} + \mathbf{c} \cdot \nabla E$ is the rate of change following the crest.      
Along the unforced free surface, 
equation (\ref{eq:Tulin}) 
reduces to 
\begin{align}
\dfrac{D_cE}{Dt} + \nabla \cdot \left(\left(\mathbf{u}-\mathbf{c}\right)E\right)=& 0 \label{eq:Tulin2}
\end{align}
which shows that $(\mathbf{u}-\mathbf{c})$ is the relevant flux velocity transporting energy to the growing crest. 

These theoretical aspects taken in context with allied observational and computational considerations, motivated our investigation of the behaviour of the local energy flux $\mathbf{F}$ in relation to the local energy density $E$ in the neighbourhood of the wave crest in the fixed frame of reference. Here equation (\ref{eq:Phillips}) can be used, along with the incompressibility condition $\nabla \cdot \mathbf{u}=0$, to quantify the ratio $\mathbf{F}/E$ as follows:  
\begin{align}
\mathbf{F}/E = \mathbf{u}\left(E+p\right)/E \label{eq:ratio}
\end{align}
On the free surface, $\mathbf{F}/E$ reduces to the surface fluid velocity $\mathbf{u}$.  Given the intrinsic relevance of the crest velocity $\mathbf{c}$ highlighted above, we adopt $\mathbf{c}$ as the natural normalising velocity for the flux ratio in equation (\ref{eq:ratio}) and introduce a breaking onset threshold parameter $\mathbf{B}$ as
\begin{align}
\mathbf{B}=\mathbf{F}/\left(E \Vert \mathbf{c} \Vert\right) \label{eq:ratio2}
\end{align}

The above analysis indicates the crest speed is a key variable in the breaking onset threshold parameter $\mathbf{B}$, as mentioned above in section \ref{sec:kinematics}.  The recent study by \citet{Banner2014} and \citet{Barthelemy2015}
provides new insights into crest speed behaviour in the unsteady evolution of nonlinear dispersive water wave packets.
Specifically, every wave in an unsteady dispersive wave group experiences a dynamic leaning cycle. 
Crests and troughs enter at the rear of the wave group first leaning forward, then transitioning through symmetry and subsequently leaning backward as they propagate towards the front of the wave group. In deep water. this leaning cycle creates a systematic crest slowdown to about $80\%$ of the linear phase velocity $c_0$, with only a weak dependence on the (local) steepness of the crest or trough. 
Here 
$c_{0}$ is the linear phase speed corresponding to the peak frequency $\omega_{0}$ of the local frequency spectrum of the wave packet, assuming the linear dispersion relation.
However, in shallower water depth, the waves are less dispersive and the crest slowdown effect reduces, with crest speeds approaching the phase speed. As will be seen below, using the crest speed $c$ underpins our key finding of a robust breaking onset threshold.   

During the crest life cycle, the local energy flux speed becomes maximal near the crest point, as will be shown below.
Since $\mathbf{F}$ and $\mathbf{c}$ are vectors, we have two convenient choices to construct their ratio: either to project along the wave propagation direction (taken here as the $x-$direction) or to use norms. This leads to the two following dimensionless quantities:
\begin{equation} 
B_{x}=F_{x}/(Ec_{x}) \quad \mbox{or} \quad B=\Vert\mathbf{F}\Vert/(E\Vert\mathbf{c}\Vert) \,. \label{eq:breakingcriterionFx}
\end{equation}
Since it is found that there is no difference between the two ratios when the crest reaches its maximum (the vertical components of the flux and of the crest speed vanish),
we only explore the validity of $B_x$ as a breaking threshold parameter that embraces wave kinematics and energetics. As discussed in detail below, the distributions of $E$, $\mathbf{F}$ and $\mathbf{B}$ provide key insights into the surface and subsurface manifestation of breaking onset.
We note that the bottom depth topography does not play an explicit role in the above discussion of our breaking onset threshold. 

\subsection{Breaking onset threshold and its surface signature}

Here we report results investigating our breaking onset criterion in terms of a parametric threshold band, as described in detail in section \ref{sec:energyfluxconsideration}. This band is bounded below by a maximal non-breaking condition, and above by a marginal breaking condition, for individual waves in the local carrier wave system. 
Fortuitously, the threshold parameter specified in equation (\ref{eq:breakingcriterionFx}) has a simple signature along the free surface.


As explained above, the pressure $p$ at the free surface is assumed constant and is taken as zero without loss of generality. The energy flux velocity then reduces to the fluid particle velocity and our proposed breaking criterion for $B_{x}$, based on excess energy flux of the marginal breaking case over the corresponding maximal recurrent case, remarkably reduces to a kinematic criterion at the free surface:
\begin{equation} \label{eq:surfacethreshold}
B_x=F_x/ Ec_x=u_x / c_x>threshold
\end{equation}
Accordingly, we performed a suite of numerical simulations to investigate the behaviour of $B_{x}$ at the free surface, as defined in equation \ref{eq:surfacethreshold}.  As described in detail below, the onset of breaking was found to occur when $0.85<B_{x}<0.86$  for our entire ensemble of numerical experiments addressing 2D and 3D, deep-water and intermediate-depth cases over flat bottom topography. In addition to this remarkable result, it is noteworthy that after factoring out the reduced crest speed in deep water ($ \approx 0.8 c_0$), this corresponds to $u/c_0 \approx 0.68$, which is well below the often-quoted classical kinematic breaking criterion $u/c_0 > 1$.  

In regard to 3D breaking validation, energy flux is a vector quantity and our criterion should be able to accommodate the additional lateral energy flux in 3D converging cases. While our 3D numerical simulations are limited to a single set (C3N5), it is a representative case with very strong lateral convergence, with a focal distance of approximately five carrier wavelengths.  

While the surface-based $B_{x}$ criterion provides the most convenient operational breaking onset threshold criterion, the associated subsurface distribution of $B_{x}$, defined in equation \ref{eq:breakingcriterionFx}, was also investigated and representative results are reported below in section \ref{sec:subsurface}. The discussion now addresses the generic behaviour of $B_{x}$ at the free surface.

\begin{figure}
\centering
\begin{tabular}{cc}
\subfloat[\label{fig:Flux-N9A0.46-NB}Paddle
amplitude $A_p/ \lambda_p=0.232$]{\includegraphics[width=0.49\textwidth]{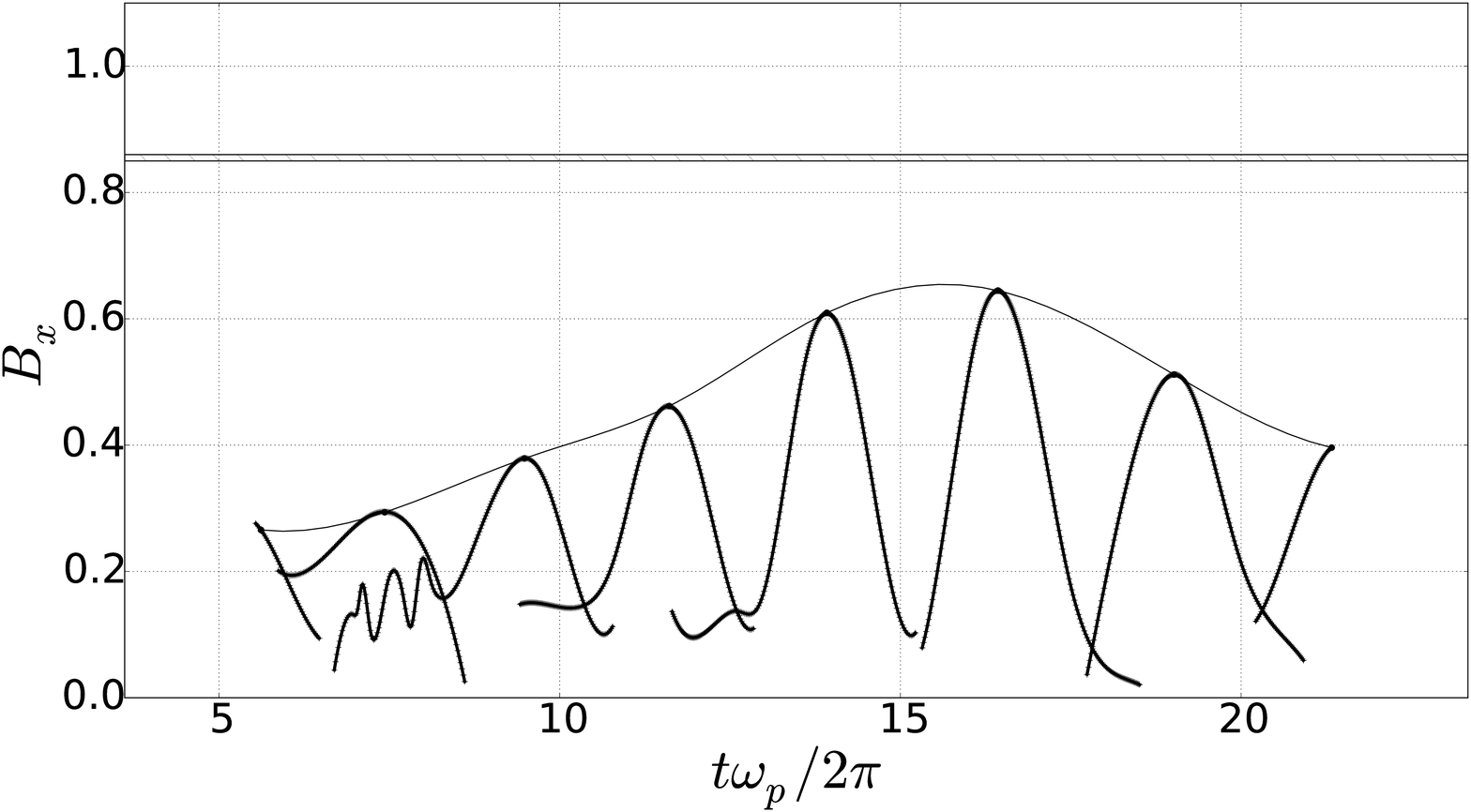}}  
&
{\subfloat[\label{fig:Flux-N9A0.47-NB}Paddle amplitude $A_p/ \lambda_p=0.237$ ]{\includegraphics[width=0.49\textwidth]{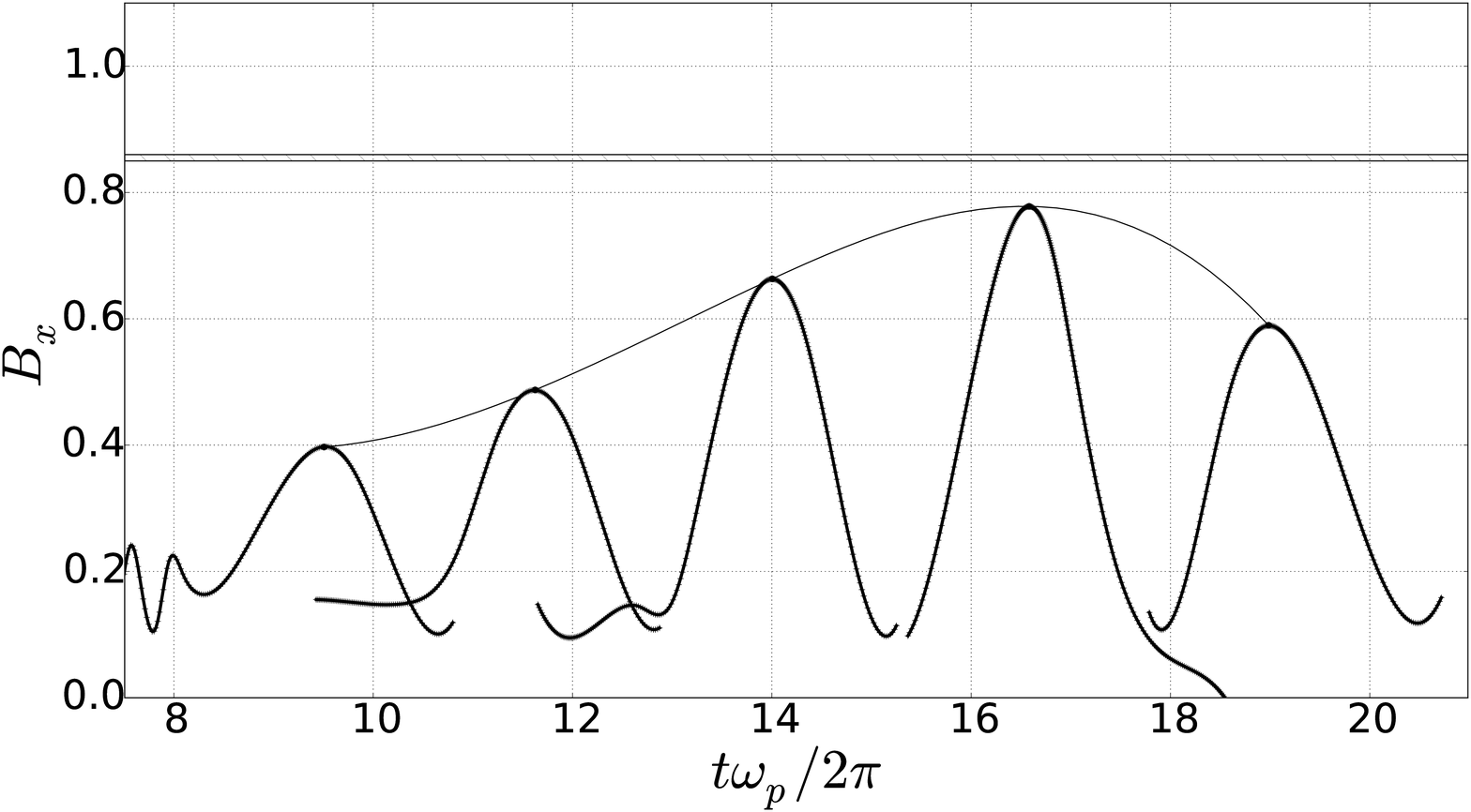}} }
\end{tabular}
\caption{Breaking criterion $B_{x}$ as a function of time for recurrent C3N9 2D chirped wave packets. Each trajectory curve shows the time evolution of the breaking parameter $B_{x}$ following each carrier wave crest maximum during the packet evolution as it grows, attains its maximum steepness without breaking and then decays. \label{fig:C3N9-NonBreaking-Flux}}
\end{figure}

\begin{figure}
\centering
\begin{tabular}{cc}
\subfloat[\label{fig:Flux-N9A0.48-B}Paddle amplitude $A_p / \lambda_p=0.242$]{\includegraphics[width=0.49\textwidth]{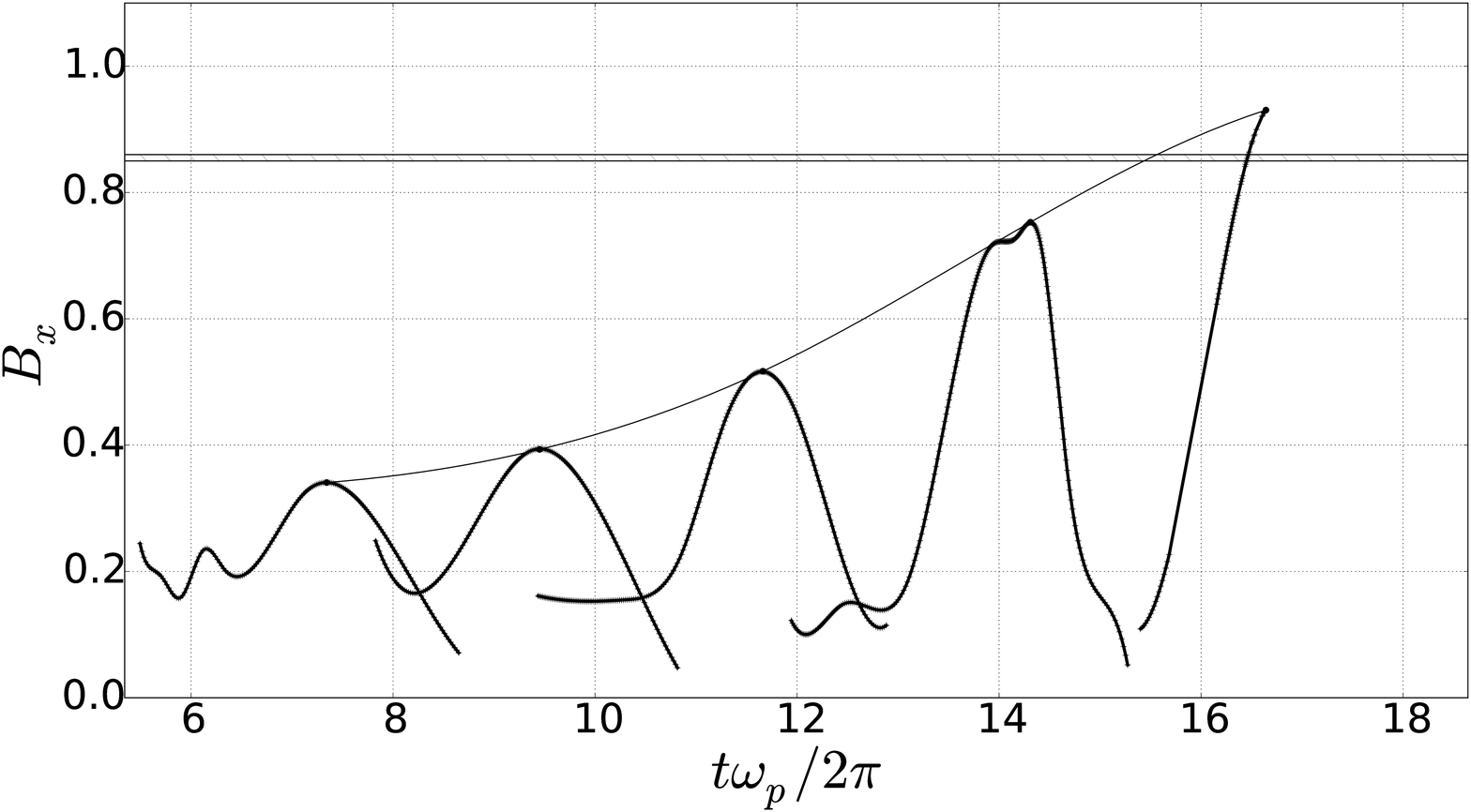}} &
\subfloat[\label{fig:Flux-N9A0.49-B}Paddle amplitude $A_p/ \lambda_p=0.247$]{\includegraphics[width=0.49\textwidth]{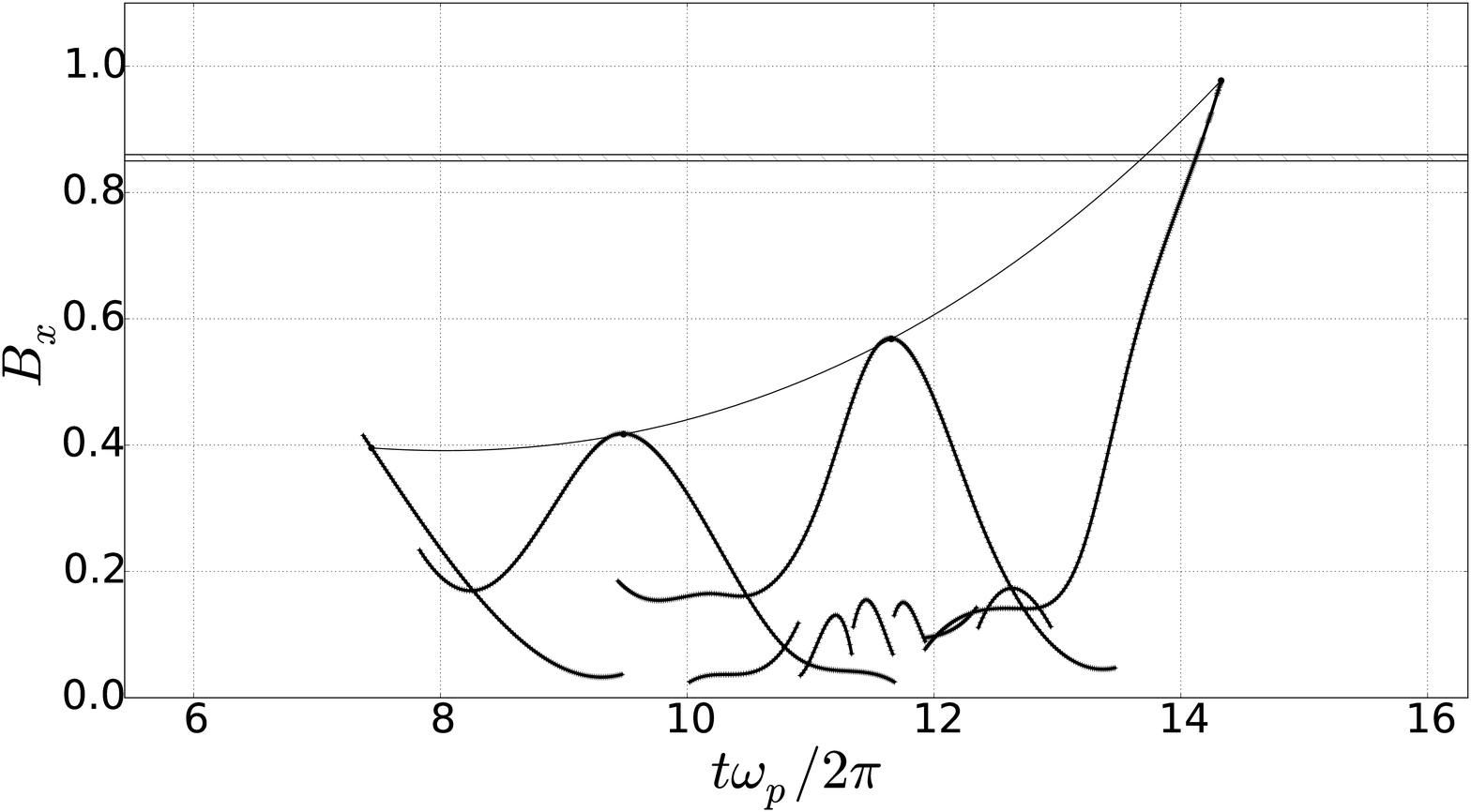}}
\end{tabular}
\caption{Breaking parameter $B_{x}$ as a function of time for breaking C3N9 2D chirped wave packets. Each trajectory curve show the time evolution, up to breaking onset, of the breaking parameter $B_{x}$ following each carrier wave crest maximum during the packet evolution. The trajectory of the breaking crest clearly crosses the proposed breaking onset threshold $[0.85,0.86]$. \label{fig:C3N9-breaking-Flux}}
\end{figure}

Representative behaviour of $B_x$ is shown in figures \ref{fig:Flux-N9A0.46-NB}, \ref{fig:Flux-N9A0.47-NB}, \ref{fig:Flux-N9A0.48-B} and \ref{fig:Flux-N9A0.49-B}. Figures \ref{fig:Flux-N9A0.47-NB} and \ref{fig:Flux-N9A0.48-B} show the time evolution of the breaking parameter $B_{x}$ following each crest maximum during the group propagation. The two cases illustrate near-maximum recurrence and marginally breaking 2D deep-water packets. The hatched zone is the identified threshold level of $0.85-0.86$ determined by our entire ensemble of simulations, above which all crests proceeded to break. 
In the near-maximum recurrence case (figure \ref{fig:Flux-N9A0.47-NB}), the trajectories never cross the hatched zone, in contrast with figure \ref{fig:Flux-N9A0.48-B}, which shows the marginal breaking case of the same wave group class, where the trajectory of $B_{x}$ clearly crosses the threshold.

In these examples, the computed spline passing through all the local crest maxima also shows their longer-term evolution trajectory. In the breaking case, this spline crosses the hatched breaking threshold zone before breaking subsequently initiates, providing advance warning of up to half a carrier wave period. This behaviour is representative of all investigated breaking cases.

Our key results for the proposed breaking onset threshold are based on $B_{x}$ derived from the ensemble of systematic numerical simulation cases shown in table \ref{tab:case}.  For each generic packet type, the results are ordered according to increasing paddle amplitude, with the transition from maximal non-breaking to marginal breaking highlighted. The C3N5 results at the top of the table confirm the insensitivity of the results to the resolution, as discussed in detail in section \ref{sec:detailCv} of the appendix. 

Figure \ref{fig:Flux-threshold}, the key figure in this paper, provides a comprehensive summary of the performance of the breaking onset threshold for the present data set. The breaking onset parameter $B_{x}$ for \emph{every} crest in each packet evolution is plotted against the corresponding crest steepness $S_{c}$. The figure shows recurrent cases at their maximum height and breaking crests at their onset. Both 2D and 3D, deep and intermediate water depth cases are included. The vertical hatched zone on the right represents the classical Stokes local steepness limit expressed in terms of $S_{c}$ rather than $ak$. The horizontal hatched zone $\left[0.85<B_{x}<0.86 \right]$ is the breaking threshold determined from our ensemble of numerical simulations. This figure highlights two significant findings. The major finding is the discovery of a clear separation between recurrent and breaking crests. For recurrence cases, our proposed breaking onset parameter $B_{x}$ is always less than $0.85$, below which crests were never found to break. However, $B_x$ is always greater than $0.86$ when breaking occurs. 
This identifies the breaking onset threshold zone for $B_{x}$ as $[0.85,0.86]$. 

The other finding concerns the relationship between crest steepness at maximum wave height and breaking onset.
Once again it is seen that crest steepness is clearly not a threshold variable that is able to discriminate between breaking and recurrence. This is evident as the local steepness for breaking crests can be lower than the local steepness of recurrent evolution cases for a given depth (or $kd$). We note that a straight line can be drawn between $(0,0)$ and $(S_{c_{max}},0.855)$. All the symbols for the individual crest are closely aligned at low steepness and then a departure from this trend occurs due to nonlinearities. The $B_{x}$ parameter then grows faster than the steepness. This transition occurs at higher steepness levels in intermediate water depth than in deep water. The limiting case for shallow water conditions ($kd\rightarrow0$) seems to converge towards the deep water Stokes limiting steepness. Table \ref{tab:case} summarises the cases processed for this study, with their properties. 

\begin{figure}
\centering %
\includegraphics[width=0.9\textwidth]{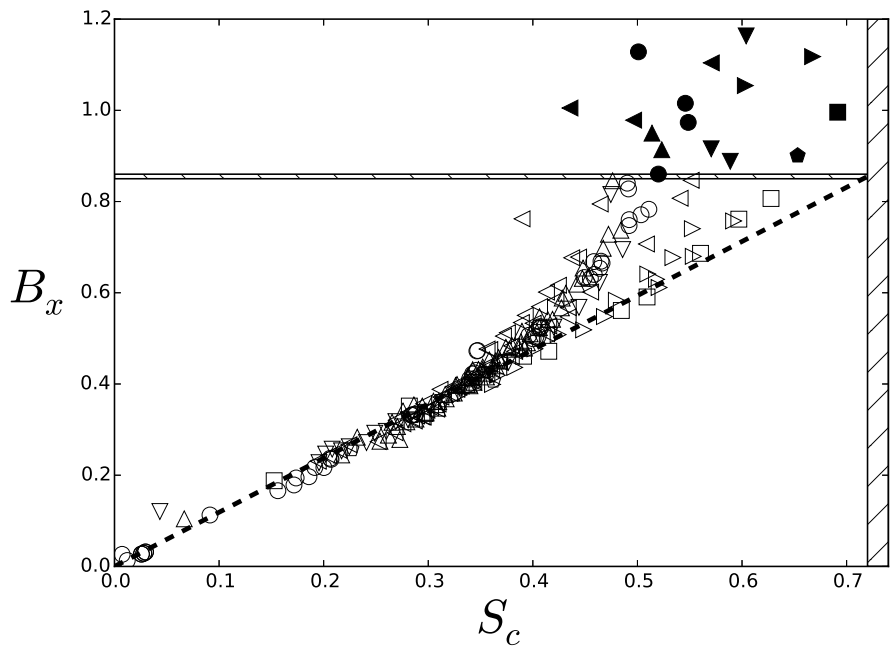} %
\caption{Breaking parameter $B_{x}$ plotted against local steepness $S_{c}$. Each point was obtained by tracking the maximum $B_x$ for every crest in each packet evolution documented in table \ref{tab:case}. The horizontal hatched zone at $0.85<B_{x}<0.86$ is the threshold which segregates breaking from non-breaking cases; the vertical hatched zone $S_{c}>0.72$ is the deep water Stokes limit. Hollow symbols represent recurrent crests and solid symbols represent breaking crests. The wave group families are labelled as follows:- circles: C3N5 2D deep water;  downward-pointing triangles: C3N5 3D deep water; upward-pointing triangles: C3N7 2D deep water; leftward-pointing triangles: C3N9 2D deep water; rightward-pointing triangles: C3N9 2D $0.2<d/ \lambda_{p}<1$; squares: C3N9 2D $d/ \lambda_{p}=0.2$; pentagons: C3N9 3D $0.2\leq d/ \lambda_{p}<1$.\label{fig:Flux-threshold}}
\end{figure}

\begin{table}
\begin{center}
\begin{tabular}{lccccccc}
Name & $d/\lambda_p$ & Nodes / $\lambda_p$ & $\mathbf{A_p / \lambda_p}$& $S_c$ Max & Breaking & $B_x$ & Figure \\
\hline
C3N5A0.05                       &$1$ & 16       &0.02521 &$0.03$ & No & $0.012$  & \\
C3N5A0.3                        &$1$ & 16       & 0.1513 &$0.206$ & No & $0.237$ & \\
C3N5A0.514                      &$1$ & 16 &0.2592 &$0.519$ & No & $0.575$ &    \\
C3N5A0.516                      &$1$ & 16 &0.2602 &$0.527$ & No & $0.606$ & \\
C3N5A0.518                      &$1$ & 16 &0.2612 &$0.533$ & No & $0.761$ &  \\
\textbf{C3N5A0.51852}                      &\textbf{1} & \textbf{16} &\textbf{0.26133} &\textbf{0.5335} & \textbf{No} & \textbf{0.831} & \ref{fig:convergence} \\
\textbf{C3N5A0.51856}                      &\textbf{1} & \textbf{16} &\textbf{0.26136} &\textbf{0.5338} & \textbf{Yes} & \textbf{0.869} & 
\ref{fig:convergence} \\
C3N5A0.519                      &$1$ & 16    &0.2617 &$0.534$ & Yes & $0.8717$ &   \\
C3N5A0.53                       &$1$ & 16    &0.2673 &$0.56$ & Yes & $1.03$ & \\
C3N5A0.56                       &$1$ & 16    &0.2824 &$0.563$ & Yes & $1.059$ & \\
\hline
C3N5A0.508                      &$1$ & 32       & 0.2562 &0.498   & No & 0.631  &   \\
\textbf{C3N5A0.511}                      &\textbf{1} & \textbf{32} &\textbf{0.2577} &\textbf{0.509} & \textbf{No} & \textbf{0.841} &  \ref{fig:convergence} \\
\textbf{C3N5A0.514}                      &\textbf{1} & \textbf{32} &\textbf{0.2592} &\textbf{0.520} & \textbf{Yes} & \textbf{0.860} & 
\ref{fig:Inside-Wave-2D},  \ref{fig:convergence}   \\
C3N5A0.516                      &$1$ & 32 &0.2602 &$0.545$ & Yes & $1.015$ & \\
C3N5A0.518                      &$1$ & 32 &0.2612 &$0.500$ & Yes & $1.128$ & \\
C3N5A0.519                      &$1$ & 32    &0.2617 &$0.466$ & Yes & $1.059$ & \\
\hline
C3N5A0.32X10                    &$1$ & 16    &0.1614 &$0.420$ & No & $0.524$ & \\
\textbf{C3N5A0.33X10}                    &\textbf{1} & \textbf{16}    &\textbf{0.1664} &\textbf{0.474} & \textbf{No} & \textbf{0.815} & \\
\textbf{C3N5A0.34X10}                   &\textbf{1} & \textbf{16}    &\textbf{0.1715} &\textbf{0.588} & \textbf{Yes} & \textbf{0.888} & \\
C3N5A0.35X10                    &$1$ & 16    &0.1765 &$0.603$ & Yes & $1.163$ & \\
C3N5A0.36X10                    &$1$ & 16    &0.1815 &$0.514$ & Yes & $0.910$ & \ref{fig:Inside-Wave-3D}\\
\hline
C3N7A0.41                       &$1$ & 16    &0.2067 &$0.344$ & No & $0.429$ & \\
C3N7A0.42                       &$1$ & 16    &0.2118 &$0.358$ & No & $0.452$ & \\
C3N7A0.43                       &$1$ & 16    &0.2168 &$0.374$ & No & $0.480$ & \\
C3N7A0.44                       &$1$ & 16    &0.2219 &$0.391$ & No & $0.509$ & \\
C3N7A0.45                       &$1$ & 16    &0.2269 &$0.406$ & No & $0.547$ & \\
C3N7A0.46                       &$1$ & 16    &0.232 &$0.428$ & No & $0.588$ & \\
C3N7A0.47                       &$1$ & 16    &0.237 &$0.448$ & No & $0.654$ & \\
\textbf{C3N7A0.48}                     &\textbf{1} & \textbf{16}    &\textbf{0.242} &\textbf{0.484} & \textbf{No} & \textbf{0.737} & \\
\textbf{C3N7A0.49}                     &\textbf{1} & \textbf{16}    &\textbf{0.2471} &\textbf{0.513} & \textbf{Yes} & \textbf{0.944} & \\
C3N7A0.50                       &$1$ & 16    &0.2521 &$0.523$ & Yes & $0.964$ & \\
\hline
C3N9A0.42                       &$1$ & 16    &0.2118 &$0.360$ & No & $0.477$ & \\
C3N9A0.43                       &$1$ & 16    &0.2168 &$0.375$ & No & $0.510$ & \\
C3N9A0.44                       &$1$ & 16    &0.2219 &$0.392$ & No & $0.550$ & \\
C3N9A0.45                       &$1$ & 16    &0.2269 &$0.413$ & No & $0.601$ & \\
C3N9A0.46                       &$1$ & 16    &0.232 &$0.437$ & No & $0.667$ & \ref{fig:Flux-N9A0.46-NB}\\
\textbf{C3N9A0.47}                       &\textbf{1} & \textbf{16}    &\textbf{0.237}  &\textbf{0.464} & \textbf{No} & \textbf{0.788} &  \ref{fig:Flux-N9A0.47-NB}\\
\textbf{C3N9A0.48}                       &\textbf{1} & \textbf{16}    &\textbf{0.242} &\textbf{0.496} & \textbf{Yes} & \textbf{0.952} &  \ref{fig:Flux-N9A0.48-B}\\
C3N9A0.49                       &$1$ & 16    &0.2471 &$0.433$ & Yes & $0.977$ &  \ref{fig:Flux-N9A0.49-B}\\
\hline
\textbf{D0.4C3N9A1.05 }                  &\textbf{0.2}   & \textbf{16}&\textbf{0.529} &\textbf{0.616} & \textbf{No} & \textbf{0.812} & \\
\textbf{D0.4C3N9A1.07}                   &\textbf{0.2}   & \textbf{16}&\textbf{0.54} &\textbf{0.682} & \textbf{Yes} & \textbf{1.001} & \\
\hline
\textbf{D0.5C3N9A0.95 }                  &\textbf{0.25}  & \textbf{16}&\textbf{0.48} &\textbf{0.595} & \textbf{No} & \textbf{0.730} & \\
\textbf{D0.5C3N9A0.97}                   &\textbf{0.25}  & \textbf{16}&\textbf{0.489} &\textbf{0.667} & \textbf{Yes} & \textbf{0.944} & \\
\textbf{D0.5C3N9A0.55X10 }               &\textbf{0.25}  & \textbf{16}&\textbf{0.277} &\textbf{0.642} & \textbf{Yes} & \textbf{0.997} & \\
\hline
\textbf{D0.75C3N9A0.77 }                 &\textbf{0.375} & \textbf{16}&\textbf{0.388} &\textbf{0.552} & \textbf{No} & \textbf{0.713} & \\
\textbf{D0.75C3N9A0.79 }                 &\textbf{0.375} & \textbf{16}&\textbf{0.398} &\textbf{0.602} & \textbf{Yes} & \textbf{0.875} & \\
\hline
\textbf{D1C3N9A0.65 }                    &\textbf{0.5}   & \textbf{16}&\textbf{0.33} &\textbf{0.541} & \textbf{No} & \textbf{0.794} & \\
\textbf{D1C3N9A0.67 }                    &\textbf{0.5}   & \textbf{16}&\textbf{0.338} &\textbf{0.590}  & \textbf{Yes} & \textbf{1.060} & \\
\end{tabular}
\end{center}
\caption{\label{tab:case} Maximum steepness $S_c$ and maximum $B_x$ for each simulated wave group within the several ensembles listed. In each group name, the first two N[I] in the group name denotes the number of waves [I] in the temporal packet, $A_p/ \lambda_p$ is the paddle amplitude and $d/ \lambda_p$ is the still water depth relative to the wavelength. Runs designated with X10 are 3D cases that focus at approximately $x/ \lambda_p=5$. The bold entries bracket the transition between the maximum recurrence and marginal breaking case 
for each group type.}


\end{table}


\subsection{Subsurface energy flux considerations}\label{sec:subsurface}

The breaking parameter $B$ is based on the local wave energy flux $\Vert\mathbf{F}\Vert$ normalised by the local wave energy density $E$ and the wave crest speed $\Vert\mathbf{c}\Vert$, defined at the maximum elevation of the crest. 
The energy density and the energy flux  are both well-behaved field quantities and reach their maximum at the crest point maximum. Because these quantities  include the potential energy, their value is defined relative to an arbitrary constant, the datum $z_0$. The total energy $E=\rho g \left(z-z_0\right) + \rho u^2/2$ may vanish locally, depending on the choice of the datum $z_0$.  Accordingly, to avoid spurious singularities of $\mathbf{B}$ and their effect on the $\mathbf{B}$ distribution arising within the flow domain, the datum level $z_0$ needs to be chosen outside  the flow domain. We  chose the datum as twice the depth of the flow domain, and verified that the subsurface $\mathbf{B}$ distributions were insensitive to $z_0$ lower than this. 

\begin{figure}
\centering 
\begin{tabular}{c}
\subfloat[\label{fig:breakingthresohold2D-3Dview-A} Colour-coded zones of the breaking parameter $B_x$ on the free surface at $t=13.85$. The blue line is the location of the slice seen in figure \ref{fig:breakingthresohold2D-2Dview-A} ]{\includegraphics[width=0.99\textwidth]{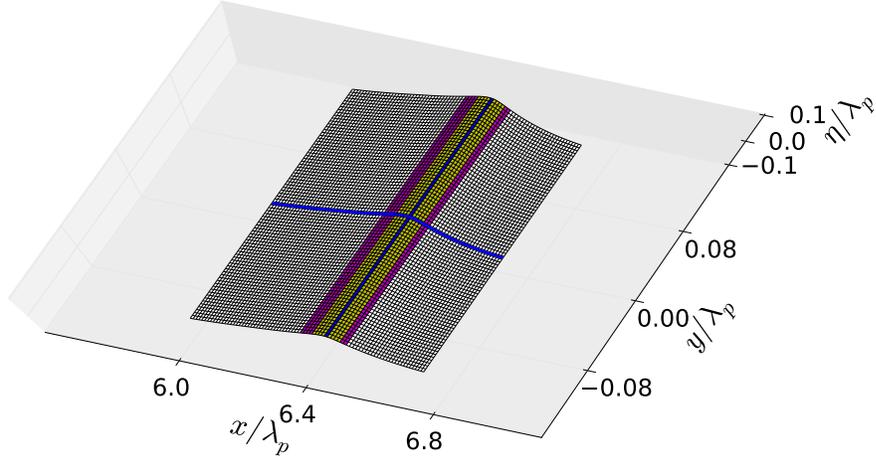}}
\\
\subfloat[\label{fig:breakingthresohold2D-3Dview-B} Same as (a) at $t=14.41$. The blue line is the location of the slice seen in figure \ref{fig:breakingthresohold2D-2Dview-B}]{\includegraphics[width=0.99\textwidth]{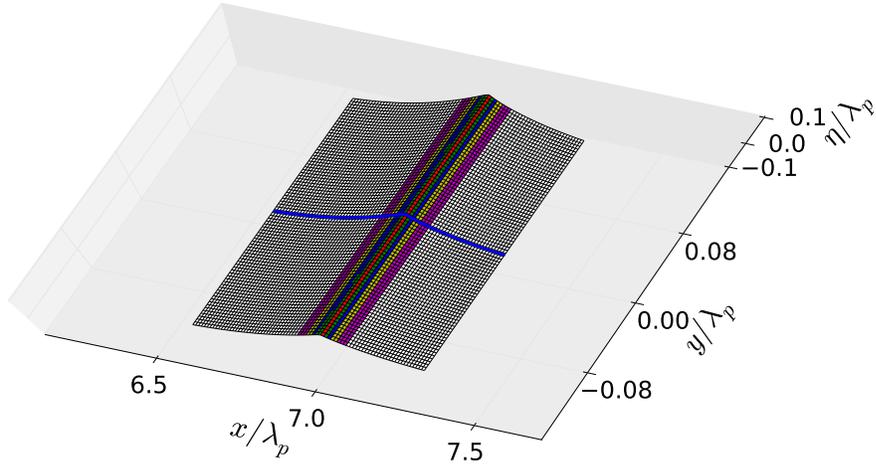}}

\\
\begin{tabular}{cc}
\subfloat[\label{fig:breakingthresohold2D-2Dview-A} Colour-coded cross-section within the slice $y=0$ shown in figure \ref{fig:breakingthresohold2D-3Dview-A} ]{\includegraphics[width=0.495\textwidth]{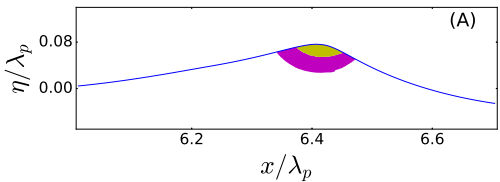}}
&
\subfloat[\label{fig:breakingthresohold2D-2Dview-B} Colour-coded cross-section within the slice $y=0$ shown in figure \ref{fig:breakingthresohold2D-3Dview-B}]{\includegraphics[width=0.495\textwidth]{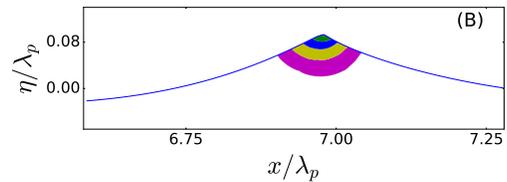}}

\end{tabular}
\end{tabular}
\caption{\label{fig:Inside-Wave-2D}Contours of the breaking criterion $B_x$ for the 2D C3N5A0.514 breaking wave case. The regions are colour-coded as follows: red: $B_x > 0.85$; green: $0.7 < B_x < 0.85$; blue: $0.6 < B_x < 0.7$; yellow: $0.5 < B_x < 0.6$; magenta: $0.4 < B_x < 0.5$; white: $ B_x < 0.4$.}
\end{figure}

\begin{figure}
\centering
\begin{tabular}{c}

\subfloat[\label{fig:breakingthresohold3D-3Dview-A} Colour-coded zones of the breaking parameter $B_x$ on the free surface at $t=12.45$. The blue line is the location of the slice seen in figure \ref{fig:breakingthresohold3D-2Dview-A} ]{\includegraphics[width=0.99\textwidth]{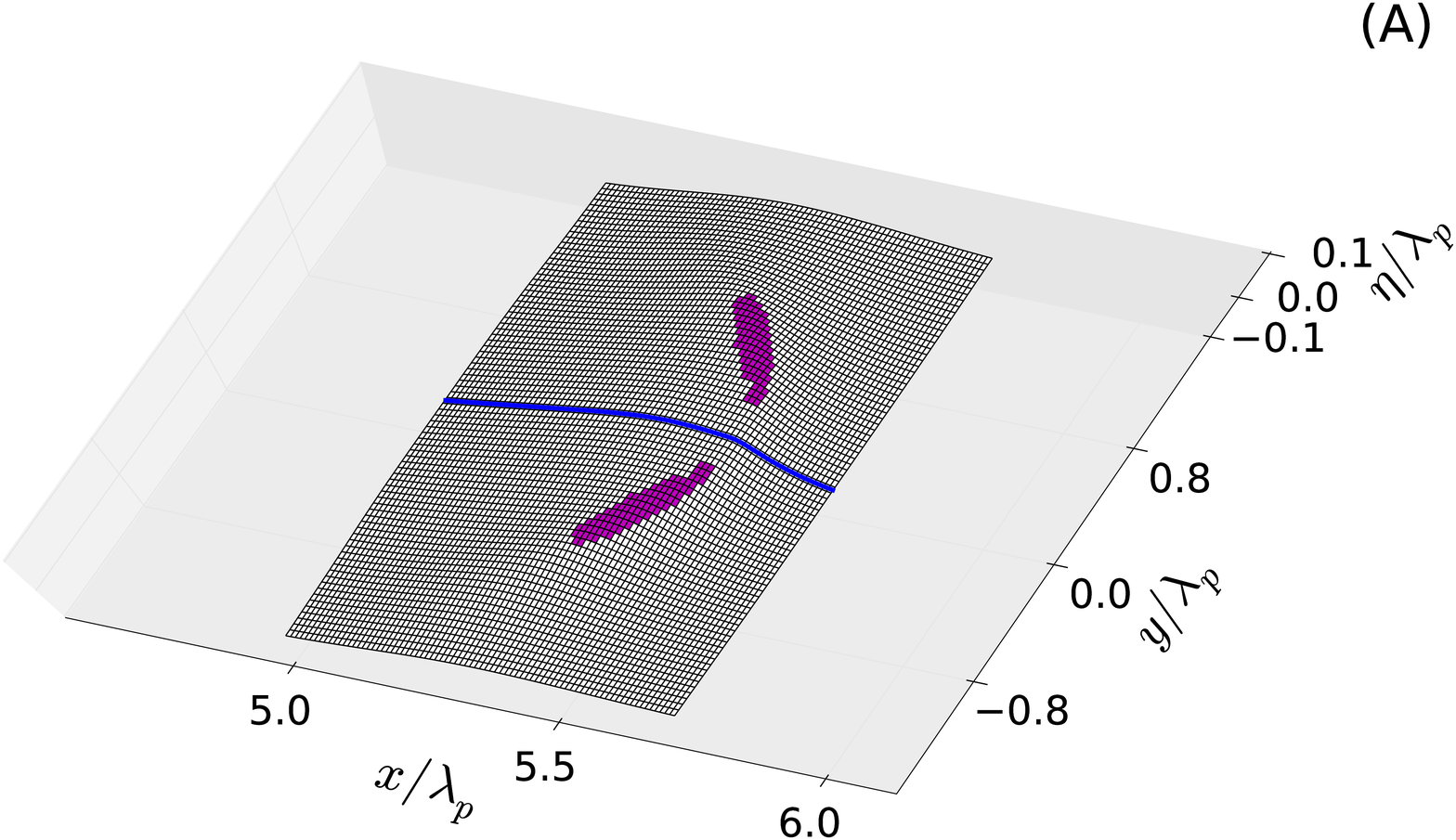}} 
\\
\subfloat[\label{fig:breakingthresohold3D-3Dview-B} Same as (a) at $t=12.84$. The blue line is the location of the slice seen in figure \ref{fig:breakingthresohold3D-2Dview-B}]{\includegraphics[width=0.99\textwidth]{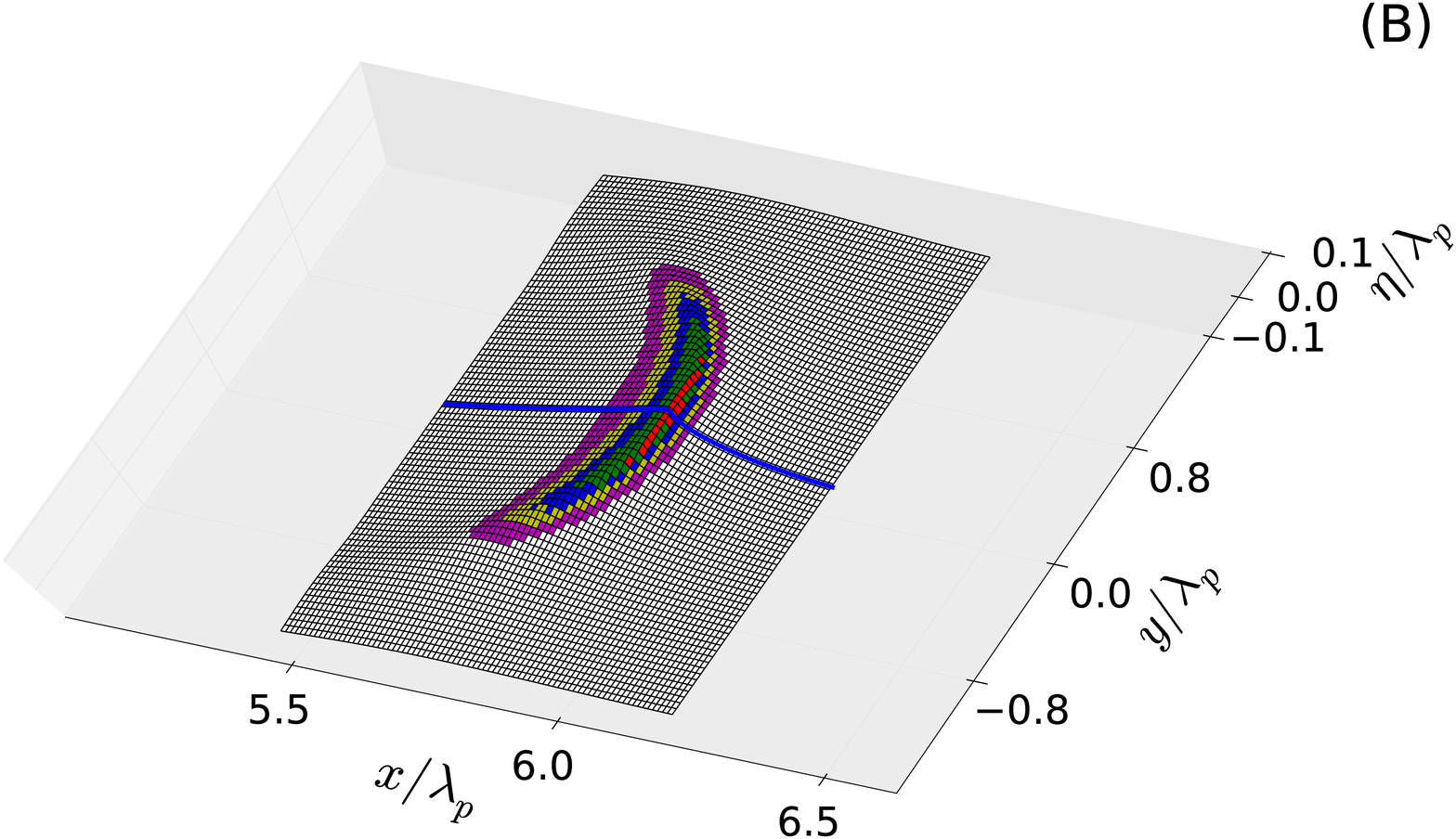}}

\\
\begin{tabular}{cc}

\subfloat[\label{fig:breakingthresohold3D-2Dview-A} Colour-coded cross-section within the slice $y=0$ shown in figure \ref{fig:breakingthresohold3D-3Dview-A}]{\includegraphics[width=0.495\textwidth]{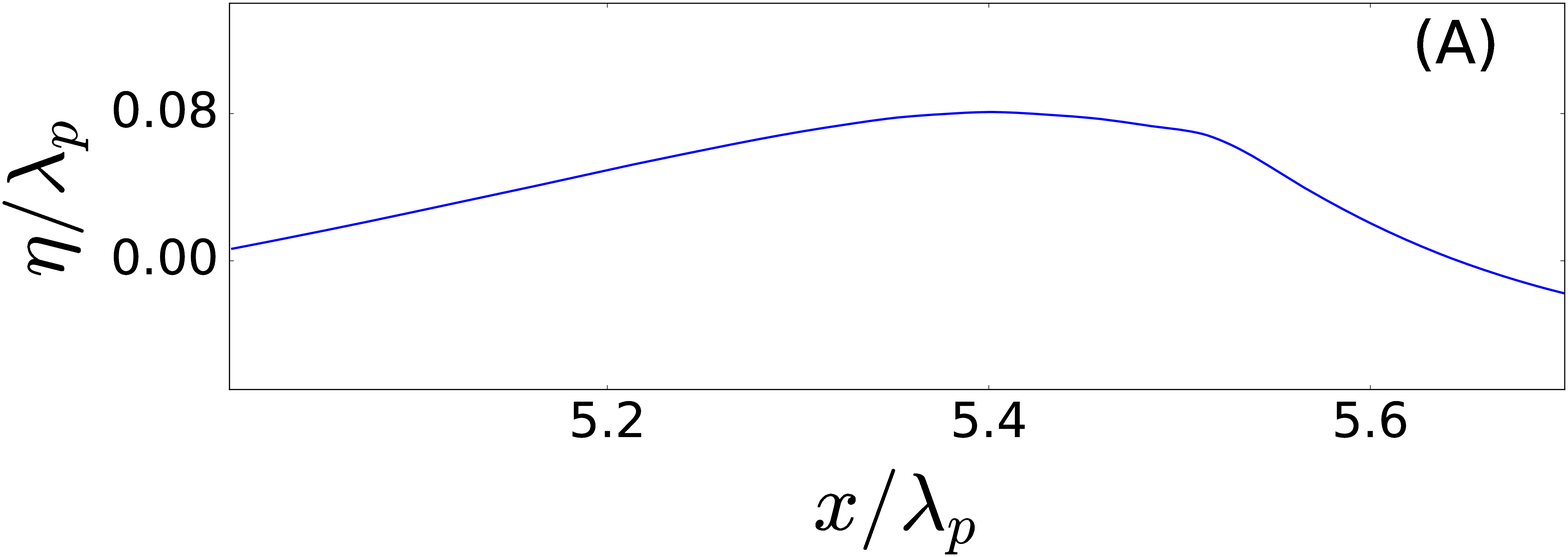}}
&
\subfloat[\label{fig:breakingthresohold3D-2Dview-B} Colour-coded cross-section within the slice $y=0$ shown in figure \ref{fig:breakingthresohold3D-3Dview-B}]{\includegraphics[width=0.495\textwidth]{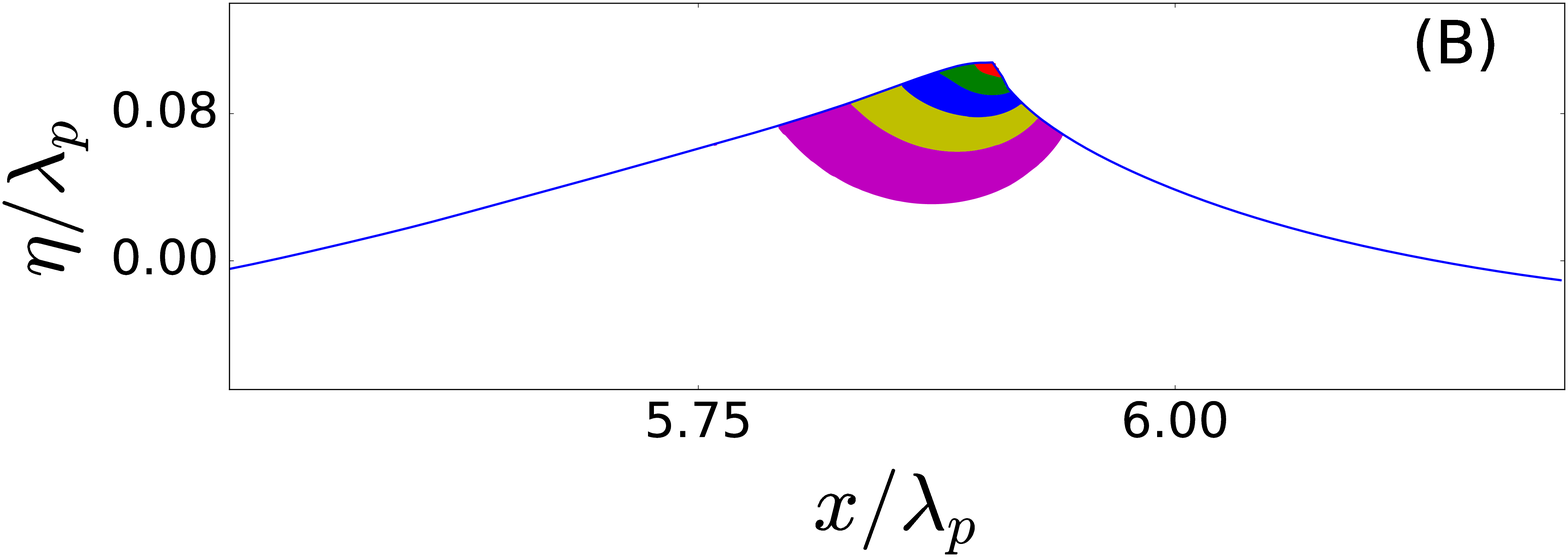}}
\end{tabular}
\end{tabular}
\caption{\label{fig:Inside-Wave-3D}
Same as figure \ref{fig:Inside-Wave-2D}, but for the 3D C3N5A0.36X10 breaking wave case.}
\end{figure}

$\mathbf{B}$ exists both on the surface and in the interior of the wave domain. It can be computed for maximum recurrent crests and for crests evolving to breaking, up to the point of breaking onset. 
Figure \ref{fig:Inside-Wave-2D} and figure \ref{fig:Inside-Wave-3D} highlight key properties of interest, respectively, of 2D and 3D waves evolving towards breaking. Each of these figures represents two snapshots (A and B) at different stages of development of the wave field. The first is a colour-coded representation showing contours of the $x-$component of the breaking onset parameter ($B_{x}$) on the
crest surface (figures \ref{fig:breakingthresohold2D-3Dview-A}, \ref{fig:breakingthresohold2D-3Dview-B}, \ref{fig:breakingthresohold3D-3Dview-A} and \ref{fig:breakingthresohold3D-3Dview-B}).
The other plot is a vertical plane slice on the symmetry axis of the crest, taken along the blue line shown in the surface plots (figures \ref{fig:breakingthresohold2D-2Dview-A}, \ref{fig:breakingthresohold2D-2Dview-B}, \ref{fig:breakingthresohold3D-2Dview-A} and \ref{fig:breakingthresohold3D-2Dview-B}).


These plots highlight the localisation of the $B_{x}$ maximum at the crest point of the tallest wave, which is similar for the 2D and 3D cases. 
 Of further interest is the spatial extent of the region involved in a breaking onset event. Based on the $B_{x}$ parameter threshold, the horizontal extent of the zone where $B_{x}$ values exceed $50\%$ of the breaking onset threshold. 

Observational validation of our new breaking criterion is challenging, due to its spatial localisation within a rapidly-evolving, compact zone. Its horizontal extent where $B_{x}$ exceeds the breaking threshold is only about $3\%$ of the wavelength. Since the non-dimensional wave crest speed is close to $0.85$ at breaking, a fixed probe will reside in this zone for only about $0.04$ non-dimensional time units. Accurate determination of the rapidly-evolving crest speed and its associated surface fluid speed is a demanding measurement. 
Careful physical experiments are needed to validate these new findings. Such studies have been undertaken and the results reported in \citet{Saket2017} for 2D waves in deep water and in \cite{Saket2017a} for intermediate water depths. 
These studies encompassed different classes of nonlinear deep water and intermediate depth wave groups, for a range of group bandwidths. The influence of wind forcing was also investigated for deep water cases.
Their results show agreement with our proposed breaking onset threshold criterion to within $2.5\%$, which is close to the experimental error bounds.

It is seen that the spatial regions where $B_{x}$ becomes appreciable in 2D and 3D breaking waves have approximately the same vertical extent, but the surface distribution of $B_{x}$ differs considerably. For 3D converging waves, the maximal values are positioned on side lobes off the symmetry axis. As the wave steepens, these lobes converge towards the symmetry axis, with $B_{x}$ maximising on the symmetry axis where the wave breaks. 
In contrast, 2D breaking waves have an almost constant spanwise distribution of $B_{x}$ values, with only minor variations found where the loci of maximum $B_{x}$ values are on either side of the symmetry axis. 

\subsection{Possible influence of surface tension on the energy fluxes}\label{sec:surf.ten}
The above investigation was carried out neglecting surface tension effects, which may become significant in zones of higher surface curvature. We assessed the validity of this assumption 
at the crest point of a 2D C3N5 wave transitioning through the $B_{x} = [0.85,0.86]$ breaking threshold with a wavelength of order $1$ m. We found the radius of curvature  $R= \left(1+\left(\zeta^{\prime}\right)^2\right)^{3/2}/\zeta^{\prime\prime} \approx 0.01$, where $z=\zeta$ specifies the free surface.  The corresponding non-dimensional water-side pressure increment $\delta p=\sigma/\rho R\approx 0.008$ is to be compared in the Bernoulli equation to the non-dimensional kinetic energy $KE=u^2/2 \approx 0.3$ and non-dimensional potential energy  $PE=g \zeta \approx 0.58$ for the datum taken at $z_0=0$. 


Thus the maximum water-side pressure increment associated with surface tension is estimated to be about $2$ orders of magnitude lower than the energies, and therefore makes negligible difference to the energy fluxes. Further, we note that crest point curvature of our numerical wave profiles as they transition the breaking onset threshold, \emph {which is well in advance of visual crest overturning}, are only about three times the observed crest curvature of $O\left(1 ~m\right)$ wavelength laboratory waves just before they proceed to overturn (e.g. see \cite{Qiao2001}, Figure 5(a)).

For water waves with even shorter wavelengths, there is the potential for our proposed breaking onset threshold to be modified by the effects of surface tension. This will only be resolved through detailed investigation.

\section{Comparison 
with previously proposed breaking onset criteria}  \label{sec:comparisonpreviouscriteria} 

Our results do not support the criterion described by \cite{Tulin2000} which links breaking onset to the crest fluid speed exceeding the linear group speed of the wave packet. 

Based on our results, for deep water waves their threshold is equivalent to  $B_{x} = 0.62$, which signals breaking onset at a considerably lower value than our $B_{x}$ threshold. However, in section \ref{sec:subsurface} we highlighted the very close agreement (within $2.5\%$) between our $B_{x}$ threshold value and the measured threshold reported by \citet{Saket2017} and \citet{Saket2017a}) for modulationally breaking deep water and intermediate water depth wave packets.
Further, as reproduced in \citet{Saket2017}, our very steep deep water recurrent waves routinely exceed the proposed group velocity-based threshold of \cite{Tulin2000} without breaking. 
For intermediate water depth waves, as the linear group speed approaches the linear phase speed, closer agreement with the \cite{Tulin2000} result might result for a specific water depth/wavelength ratio. 

\citet{Perlin2013} cite two studies that carefully compare the highest water speeds with coincident wave speeds. Just prior to overturning, \cite{Perlin1996} found a ratio of maximum water speed to linear wave speed of $0.74$ for a single case of deep water plunging breaking. For intermediate water depths, \cite{Chang1998} found a ratio of maximum water speed to linear wave speed of $0.86$ prior to breaking. However, as shown by \cite{Banner2014}, there is a systematic slowdown of the crest point as it transitions through its local maximum and this crest slowdown was not taken into account by these investigators.
After the crest slowdown is taken into account, their measurements are consistent with our findings, as both investigations reported water speeds exceeding $0.85$ of the crest speed at breaking onset, with subsequent water speeds exceeding wave speeds during the overturning breaking process.

\citet{Saket2017} summarise results from a suite of other investigations that compared crest water speed with the crest speed just prior to breaking onset. The most important of these studies is the work of \cite{Stansell2002}, whose findings were found to be consistent with \citet{Saket2017}, once plausible and appropriate corrections were made to the near-surface velocity structure reported by \cite{Stansell2002}.

\section{Discussion and conclusions} \label{sec:conclusion}
A new breaking onset threshold criterion based on energy flux considerations has been developed for water waves and its applicability investigated for 2D and 3D chirped focusing wave groups in deep and intermediate water depths. While the energy flux that initiates breaking arises from energy focusing within the whole wave group, we found that the initial instability occurs within a very compact region, about 3 percent of the wavelength in horizontal extent, centred on the maximum wave crest. 

We found that depth-integrated quantities were not able to detect the signature of breaking generically in the phase-resolved wave motion. A more localised examination of the flow near the free surface was necessary to detect the initiation of breaking and to find a consistent deterministic indicator. Our new breaking criterion is based on the strength of the local energy flux relative to the local energy density, normalised by the local crest speed. This non-dimensional parameter $\mathbf{B}$, which reduces to $B_{x}$ at the crest point (here $x$ is the direction of propagation), simplifies fortuitously for zero surface pressure. On the free surface, it reduces to a kinematic condition for the ratio of crest fluid speed to crest point speed. 

For the condition of zero surface pressure forcing, our suite of numerical experiments using a significant range of chirped wave packets has shown that our new dynamically-based breaking onset parameter $B_{x}$ has a generic threshold value computed to be in the range $\left[0.85, 0.86 \right]$, above which any local crest will undergo breaking onset. This threshold band is applicable everywhere in the fluid domain 
and has important new consequences.

Surface projection of our criterion is needed to make comparisons with previously proposed kinematic criteria. When projected along the surface, our criterion predicts breaking onset considerably in advance of the kinematic breaking criterion, where breaking is associated with crest fluid speed exceeding the crest point speed. For maximally steep non-breaking waves, the crest fluid speed cannot exceed $0.85$  of the crest speed.  When referenced to the linear wave speed $c_0$, after factoring in the generic crest slowdown for focusing dispersive deep water wave packets for which $c\approx 0.8 c_0$ (see sections \ref{sec:kinematics} and \ref{sub:Breaking-criterion-based}), our $B_{x}$ threshold predicts breaking onset for deep water waves when the crest fluid speed to linear phase speed ratio $u/c_0$ exceeds $0.68$. In progressively shallower water depth conditions, the generic crest slowdown reduces ($c \rightarrow c_0$) and $u/c_0$ increases towards $\left[0.85, 0.86 \right]$. In any event, it is evident that our breaking onset threshold band, which follows from energy flux considerations, occurs well before the water speed outruns the crest speed.

Overall, remarkable robustness of this new breaking onset threshold was found for the diverse range of 2D and 3D chirped focusing wave packets on deep and intermediate depth flat bottom topography investigated in this study. The threshold is also consistent with the findings of available comparable experiments for wavelengths of $O\left(1 m\right)$  (\cite{Saket2017} and \cite{Saket2017a}). Further investigation is needed for shorter waves when surface tension effects may become important and also for other commonly encountered breaking wave scenarios, such as waves shoaling on inclined bottom topography, coalescing wave groups and waves on uniform and shear currents. For these scenarios, since there is no explicit dependence on water depth, mean current or energy convergence rate in our energy flux considerations (section \ref{sec:energyfluxconsideration}), we anticipate that our breaking onset threshold will be applicable to such cases. 

\section{Acknowledgements}
Funding for this investigation was provided by the Australian Research Council under Discovery Project DP120101701. Also, MB acknowledges ongoing support from the US Office of Naval Research during this project. FD acknowledges partial support by the European Research Council (ERC) under the research project ERC-2011-AdG 290562-MULTIWAVE and Science Foundation Ireland under grant number SFI/12/ERC/E2227

\section{Authors' contributions}

XB performed all aspects of the computations. MB coordinated the scientific effort in close collaboration with XB, FF, WP, FD and MA. XB and MB pursued the forensics underpinning the discovery. They drafted this paper, with significant technical and intellectual input on the analysis and interpretation of the results from WP, FF, FD and MA. 

\appendix

\section{Details of the numerical wave tank}
\subsection{The numerical wave tank} \label{sec:NWT}

\begin{figure}
\centering
\psfrag{X}[][]  { $\displaystyle \mathbf{x}$}
\psfrag{Y}[][]  { $\displaystyle \mathbf{y}$}
\psfrag{Z}[][]  { $\displaystyle \mathbf{z}$}
\psfrag{xt}[][]  { \quad $\displaystyle \mathbf{x(t)}$}
\psfrag{n}[][]  { $\displaystyle \vec{\mathbf{n}}$}
\psfrag{s}[][] { $\displaystyle \vec{\mathbf{s}}$}
\psfrag{m}[][] { $\displaystyle \vec{\mathbf{m}}$}
\psfrag{Gfs}[][] { $\displaystyle \mathbf{\Gamma_{FS}}$}
\psfrag{O}[][] { $\displaystyle \mathbf{O}$}
\psfrag{Beach}[][] { \textbf{Absorbing Piston}}
\psfrag{Gamma}[][] { $\displaystyle \mathbf{\Gamma}$}
\psfrag{Paddle}[][] {\textbf{Paddle}}
\includegraphics[height=6cm]{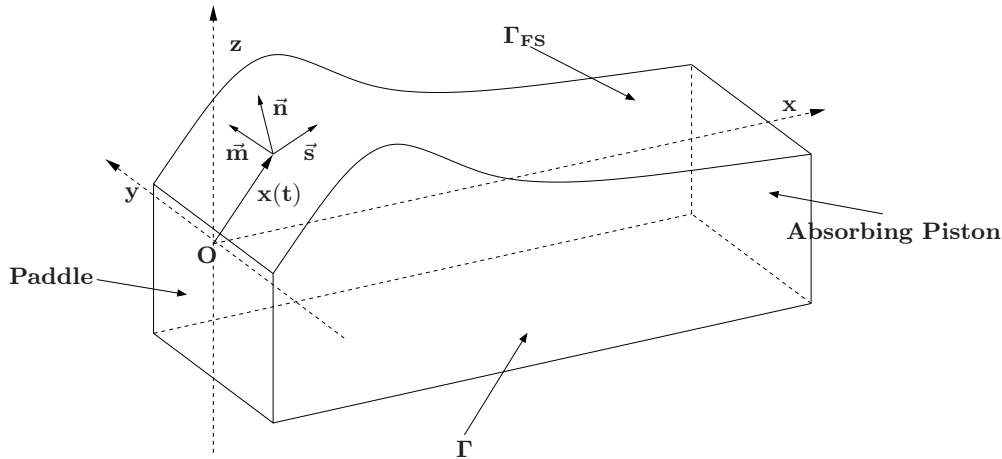}
\caption{Schematic (not to scale) of the WSIM simulation domain and nomenclature.}
\label{fig:Simulation-domain}
\end{figure}

WSIM uses a mixed Eulerian-Lagrangian time-updating scheme for irrotational motion described by the velocity potential $\phi\left(\mathbf{x},t\right)$, in a Cartesian coordinate system $\mathbf{x}=\left(x,y,z\right)$ with constant pressure at the open water surface. $z$ is the vertical upward direction and $z=0$ the still water surface (figure \ref{fig:Simulation-domain}).

Fluid velocity is defined as $\mathbf{u}=\mathbf{\nabla\phi}=\left(u,v,w\right)$. The continuity equation leads to the Laplace equation for the potential within the fluid domain $\Omega\left(t\right)$ (\ref{eq:laplace}). The symbols $\Gamma$ and $\Gamma_{FS}$ are used to denote the entire domain boundary and the free surface respectively. The equations read:

\begin{eqnarray}
\nabla^{2}\phi & = & 0 \textrm{, on }\Omega\label{eq:laplace}\\
\frac{D\mathbf{r}}{Dt} & = & \mathbf{u}=\mathbf{\nabla\phi}  \textrm{, on }\Gamma_{FS}\label{eq:kinematic}\\
\frac{D\phi}{Dt} & = & -gz+\frac{1}{2}\nabla\phi\nabla\phi-\frac{p_{0}}{\rho} \textrm{, on }\Gamma_{FS}\label{eq:dynamic}\\
\partial_{n}\phi & = & 0 \textrm{, on }\Gamma\backslash\Gamma_{FS}\label{eq:boundary}
\end{eqnarray}
where $\mathbf{r}$ is the position vector of a fluid particle on the free surface, $g$ the acceleration due to gravity, $p_{0}$ the atmospheric pressure, $\rho$ the fluid density and $\frac{D}{Dt}\left(=\frac{\partial}{\partial t}+\nabla\phi\cdot\nabla\right)$  the Lagrangian (or material) time derivative.

The free surface (domain $\Gamma_{FS}$) is described by fully-nonlinear kinematic (\ref{eq:kinematic}) and dynamic (\ref{eq:dynamic}) equations. Recent developments have implemented a 3D snake wave paddle at one vertical face of the domain which is described subsequently. The far face of the numerical wave tank from the paddle has an absorbing beach which damps any incident wave energy as described in \citet{Grilli1997}. All remaining faces of the domain have a zero-flux boundary condition ($\Gamma\backslash\Gamma_{FS}$) (\ref{eq:boundary}).
The depth of the NWT domain is equal to the wavelength ($\lambda_p$) and its size (length, width) is (12.5$\lambda_p$, 1/4$\lambda_p$) for 2D and (10$\lambda_p$, 7$\lambda_p$) for 3D computations.

\subsection{Post-processing}

\label{interiorfields}

A significant challenge within this investigation was determining quantities immediately below the highly-curved free surface. 
While the surface velocity distribution is available directly from the model output, reconstructing the subsurface velocity field values relies on the ability to correctly estimate Green's integral over the entire domain. Three complementary methods are used depending on the distance between the inner point and the boundaries.
When the distance is large enough (relative to the element size), the integration on the individual element is carried out by a Riemann quadrature on a classical Gauss-Lobatto point distribution. 
When the distance becomes small, this classical quadrature method does not retain enough precision, and the \citet{Telles1987} method is used. The Telles method consists of binary subdivisions of the integration space to maximize precision where the singularity begins to manifest itself. The precision of this method is acceptable at moderate distance from the boundary, as described in \citet{Grilli1994}.
However, the Telles method becomes inefficient when the near-boundary singularities are too strong, specifically adjacent to the highly curved surface of steep waves. Consequently, a third method was developed and implemented called Projection and Angular and Radial Transformation (PART) (see  \citet{Hayami1991}, \citet{Hayami1994}, \citet{Hayami2005a}). 

The crest and trough locations are tracked for each carrier wave in the packet  using a two-stage detection algorithm. First, extrema and the zero-crossing positions are computed semi-analytically from the same $3^{\text{rd}}$ order spline polynomials used in the BEM code. These positions are then linked together between time steps to form the space-time characteristics of the motion, where crests, troughs and zero-crossings were detected and followed in the inertial frame of reference of the wavetank. The speed of each crest was then computed as the first derivative of these trajectories in space and time.

\subsection{Details on the convergence of the NWT for marginally breaking waves}\label{sec:detailCv}

\begin{figure}
\centering
\includegraphics[width=0.85\textwidth]{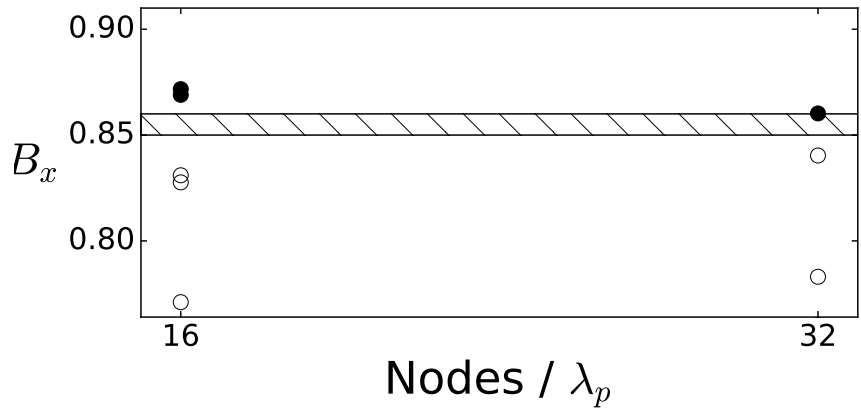}
\caption{Breaking index $B_x$ versus average number of nodes per wavelength $\lambda_p$ for the same C3N5 case. Filled circles indicate breaking crests, while open circles are recurrent crests. Each point is derived from the maximum of each individually-tracked wave as in figures \ref{fig:C3N9-NonBreaking-Flux}, \ref{fig:C3N9-breaking-Flux} and \ref{fig:Flux-threshold}.  The hatched breaking threshold zone based on our entire ensemble of derived $B_x$ values is seen to be relatively insensitive to the resolution. \label{fig:convergence}}

\end{figure}

In this section, a more detailed analysis of maximal recurrent and marginal breaking cases is made to investigate possible sensitivity of  the breaking onset parameter computed using two different resolutions: one using an average of $16$ nodes per wavelength (R16) and a higher resolution using $32$ nodes per wavelength (R32). Small differences between the runs are reported and discussed below, but are of minor consequence and the NWT is found to be sufficiently accurate at R16 resolution to objectively resolve the differences between maximum recurrent and marginal breaking cases.
For this purpose we adopted the occurrence of a vertical segment on the forward face as a consistent post-breaking onset reference for comparing  sensitivity of the paddle amplitudes, space-time locations, wave steepness and fluid velocities for the 2D C3N5 marginal breaking case computed at different resolutions. Results for our proposed breaking parameter $B_x$ are shown in figure \ref{fig:convergence}.


The numerical experiments we ran for 2D C3N5 wave packets for different resolutions showed some minor differences. 
In this instance, the paddle amplitude necessary to reach maximum recurrence or marginal breaking reduces slightly between R16 and R32 cases. 
The marginal breaking results for the slightly steeper R32 run produced slightly larger breaker amplitude, steepness and energy flux than the R16 case, consistent with the observed slightly shorter breaking onset epoch and corresponding fetch. 
However, for our primary goal of computing the derived breaking parameter $B_x$, different resolutions for the same C3N5 wave group have only a minor influence on the determination of the threshold. Figure \ref{fig:convergence} confirms the convergence of the NWT simulations in determining the threshold zone for $B_x$.  Our proposed nominal breaking threshold value $B_x=0.85$ is well-bracketed for each resolution:  $B_x$ is insensitive to the resolution, and R16 is seen to be sufficient for our purpose of establishing a discriminating breaking criterion for 2D and 3D deep and intermediate depth water focussing wave packets.

\bibliographystyle{jfm}
\bibliography{biblio-full}

\begin{thebibliography}{60}
\expandafter\ifx\csname natexlab\endcsname\relax\def\natexlab#1{#1}\fi

\bibitem[Allis(2013)]{Allis2013}
{\sc Allis, Michael} 2013 The speed, breaking onset and energy dissipation of
  3d deep-water waves. PhD thesis, Water Research Laboratory Faculty of
  Engineering UNSW.

\bibitem[Baker {\em et~al.\/}(1982)Baker, Meiron \& Orszag]{Baker1982}
{\sc Baker, Gregory~R., Meiron, Daniel~I. \& Orszag, Steven~A.} 1982
  Generalized vortex methods for free-surface flow problems. {\em Journal of
  Fluid Mechanics\/} {\bf 123}, 477--501.

\bibitem[Banner {\em et~al.\/}(2014)Banner, Barthelemy, Fedele, Allis,
  Benetazzo, Dias \& Peirson]{Banner2014}
{\sc Banner, M.\, L., Barthelemy, X., Fedele, F., Allis, M., Benetazzo, A.,
  Dias, F. \& Peirson, W.\, L.} 2014 Linking reduced breaking crest speeds to
  unsteady nonlinear water wave group behavior. {\em Phys. Rev. Lett.\/} {\bf
  112}, 114502.

\bibitem[Banner \& Peirson(2007)]{Banner2007}
{\sc Banner, Michael~L. \& Peirson, William~L.} 2007 Wave breaking onset and
  strength for two-dimensional deep-water wave groups. {\em Journal of Fluid
  Mechanics\/} {\bf 585}, 93--115.

\bibitem[Banner \& Tian(1998)]{Banner1998}
{\sc Banner, Michael~L. \& Tian, Xin} 1998 On the determination of the onset of
  breaking for modulating surface gravity water waves. {\em Journal of Fluid
  Mechanics\/} {\bf 367}~(-1), 107--137.

\bibitem[Barthelemy {\em et~al.\/}(2015)Barthelemy, Banner, Peirson, Dias \&
  Allis]{Barthelemy2015}
{\sc Barthelemy, X., Banner, M.L., Peirson, W.L., Dias, F. \& Allis, M.} 2015
  On the local properties of highly nonlinear unsteady gravity water waves.
  part 1. slowdown, kinematics and energetics. {\em arXiv:1508.06001\/} .

\bibitem[Bateman {\em et~al.\/}(2001)Bateman, Swan \& Taylor]{Bateman2001}
{\sc Bateman, W.J.D., Swan, C. \& Taylor, P.H.} 2001 On the efficient numerical
  simulation of directionally spread surface water waves. {\em Journal of
  Computational Physics\/} {\bf 174}~(1), 277 -- 305.

\bibitem[Chalikov \& Babanin(2012)]{Chalikov2012}
{\sc Chalikov, Dmitry \& Babanin, Alexander~V.} 2012 Simulation of wave
  breaking in one-dimensional spectral environment. {\em J. Phys. Oceanogr.\/}
  {\bf 42}~(11), 1745--1761.

\bibitem[Chang \& Liu(1998)]{Chang1998}
{\sc Chang, Kuang-An \& Liu, Philip L.-F.} 1998 Velocity, acceleration and
  vorticity under a breaking wave. {\em Physics of Fluids\/} {\bf 10}~(1),
  327--329.

\bibitem[Clamond \& Grue(2001)]{Clamond2001}
{\sc Clamond, Didier \& Grue, John} 2001 A fast method for fully nonlinear
  water-wave computations. {\em Journal of Fluid Mechanics\/} {\bf 447},
  337--355.

\bibitem[Craig \& Sulem(1993)]{Craig1993}
{\sc Craig, W. \& Sulem, C.} 1993 Numerical simulation of gravity waves. {\em
  Journal of Computational Physics\/} {\bf 108}~(1), 73 -- 83.

\bibitem[Dalrymple(1989)]{Dalrymple1989}
{\sc Dalrymple, R.~A.} 1989 Directional wavemaker theory with sidewall
  reflection. {\em Journal of Hydraulic Research\/} {\bf 27}~(1), 23--34.

\bibitem[Dalrymple \& Kirby(1988)]{Dalrymple1988}
{\sc Dalrymple, Robert~A. \& Kirby, James~T.} 1988 Models for very wide-angle
  water waves and wave diffraction. {\em Journal of Fluid Mechanics Digital
  Archive\/} {\bf 192}~(-1), 33--50.

\bibitem[Derakhti \& Kirby(2016)]{Derakhti2016}
{\sc Derakhti, Morteza \& Kirby, James~T.} 2016 Breaking-onset, energy and
  momentum flux in unsteady focused wave packets. {\em Journal of Fluid
  Mechanics\/} {\bf 790}, 553--581.

\bibitem[Dommermuth \& Yue(1987)]{Dommermuth1987}
{\sc Dommermuth, Douglas~G. \& Yue, Dick K.~P.} 1987 A high-order spectral
  method for the study of nonlinear gravity waves. {\em Journal of Fluid
  Mechanics\/} {\bf 184}, 267--288.

\bibitem[Ducrozet {\em et~al.\/}(2012)Ducrozet, Bonnefoy, {Le Touze} \&
  Ferrant]{Ducrozet2011}
{\sc Ducrozet, Guillaume, Bonnefoy, Felicien, {Le Touze}, David \& Ferrant,
  Pierre} 2012 A modified high-order spectral method for wavemaker modeling in
  a numerical wave tank. {\em European Journal of Mechanics - B/Fluids\/} {\bf
  34}, 19--34.

\bibitem[Duncan(2001)]{Duncan2001}
{\sc Duncan, James~H.} 2001 Spilling breakers. {\em Annual Review of Fluid
  Mechanics\/} {\bf 33}~(1), 519--547.

\bibitem[Fedele(2014)]{Fedele2014}
{\sc Fedele, Francesco} 2014 Geometric phases of water waves. {\em EPL
  (Europhysics Letters)\/} {\bf 107}~(6), 69001.

\bibitem[Fedele {\em et~al.\/}(2016)Fedele, Brennan, {Ponce de Le\'on}, Dudley
  \& Dias]{Fedele2016a}
{\sc Fedele, Francesco, Brennan, Joseph, {Ponce de Le\'on}, Sonia, Dudley, John
  \& Dias, Fr\'ed\'eric} 2016 Real world ocean rogue waves explained without
  the modulational instability. {\em Scientific Reports\/} {\bf 6}, 27715--.

\bibitem[Fochesato(2004)]{Fochesato2004}
{\sc Fochesato, Christophe} 2004 Mod{\`e}les num{\'e}riques pour les vagues et
  les ondes internes. PhD thesis, CMLA / Ecole Normale Superieure de CACHAN.

\bibitem[Fochesato \& Dias(2006)]{Fochesato2006}
{\sc Fochesato, Christophe \& Dias, Fr\'ed\'eric} 2006 A fast method for
  nonlinear three-dimensional free-surface waves. {\em Proceedings of the Royal
  Society A: Mathematical, Physical and Engineering Science\/} {\bf
  462}~(2073), 2715--2735.

\bibitem[Fochesato {\em et~al.\/}(2007)Fochesato, Grilli \&
  Dias]{Fochesato2007}
{\sc Fochesato, Christophe, Grilli, St{\'e}phan \& Dias, Fr{\'e}d{\'e}ric} 2007
  Numerical modeling of extreme rogue waves generated by directional energy
  focusing. {\em Wave Motion\/} {\bf 44}~(5), 395 -- 416.

\bibitem[Fructus {\em et~al.\/}(2005)Fructus, Clamond, Grue \&
  Kristiansen]{Fructus2005}
{\sc Fructus, Dorian, Clamond, Didier, Grue, John \& Kristiansen, Oyvind} 2005
  An efficient model for three-dimensional surface wave simulations: Part i:
  Free space problems. {\em Journal of Computational Physics\/} {\bf 205}~(2),
  665 -- 685.

\bibitem[Grilli \& Svendsen(1990)]{Grilli1990}
{\sc Grilli, S.T. \& Svendsen, I.A.} 1990 Corner problems and global accuracy
  in the boundary element solution of nonlinear wave flows. {\em Engineering
  Analysis with Boundary Elements\/} {\bf 7}~(4), 178 -- 195.

\bibitem[Grilli {\em et~al.\/}(2001)Grilli, Guyenne \& Dias]{Grilli2001}
{\sc Grilli, St{\'e}phan~T., Guyenne, Philippe \& Dias, Fr{\'e}d{\'e}ric} 2001
  A fully non-linear model for three-dimensional overturning waves over an
  arbitrary bottom. {\em International Journal for Numerical Methods in
  Fluids\/} {\bf 35}~(7), 829--867.

\bibitem[Grilli \& Horrillo(1997)]{Grilli1997}
{\sc Grilli, St{\'e}phan~T. \& Horrillo, Juan} 1997 Numerical generation and
  absorption of fully nonlinear periodic waves. {\em Journal of Engineering
  Mechanics\/} {\bf 123}~(10), 1060--1069.

\bibitem[Grilli {\em et~al.\/}(1989)Grilli, Skourup \& Svendsen]{Grilli1989}
{\sc Grilli, S.~T., Skourup, J. \& Svendsen, I.~A.} 1989 An efficient boundary
  element method for nonlinear water waves. {\em Engineering Analysis with
  Boundary Elements\/} {\bf 6}~(2), 97 -- 107.

\bibitem[Grilli \& Subramanya(1994)]{Grilli1994}
{\sc Grilli, St{\'e}phan~T. \& Subramanya, Ravishankar} 1994 Quasi-singular
  integrals in the modeling of nonlinear water waves in shallow water. {\em
  Engineering Analysis with Boundary Elements\/} {\bf 13}~(2), 181 -- 191.

\bibitem[Grilli \& Subramanya(1996)]{Grilli1996}
{\sc Grilli, S.~T. \& Subramanya, R.} 1996 Numerical modeling of wave breaking
  induced by fixed or moving boundaries. {\em Computational Mechanics\/} {\bf
  17}~(6), 374 -- 391.

\bibitem[Guyenne \& Grilli(2006)]{Guyenne2006}
{\sc Guyenne, P. \& Grilli, S.~T.} 2006 Numerical study of three-dimensional
  overturning waves in shallow water. {\em Journal of Fluid Mechanics\/} {\bf
  547}, 361--388.

\bibitem[Hayami(1991)]{Hayami1991}
{\sc Hayami, K} 1991 A projection transformation method for nearly singular
  surface boundary element integrals. PhD thesis, Computational mechanics
  institute, Wessex institute of technology, Southampton.

\bibitem[Hayami(2005)]{Hayami2005a}
{\sc Hayami, K} 2005 Variable transformations for nearly singular integrals in
  the boundary element method. {\em ublications of Research Institute for
  Mathematical Sciences\/} {\bf 41}, 821--842.

\bibitem[Hayami \& Matsumoto(1994)]{Hayami1994}
{\sc Hayami, Ken \& Matsumoto, Hideki} 1994 A numerical quadrature for nearly
  singular boundary element integrals. {\em Engineering Analysis with Boundary
  Elements\/} {\bf 13}~(2), 143 -- 154.

\bibitem[Holthuijsen \& Herbers(1986)]{Holthuijsen1986}
{\sc Holthuijsen, L.~H. \& Herbers, T. H.~C.} 1986 Statistics of breaking waves
  observed as whitecaps in the open sea. {\em J. Phys. Oceanogr.\/} {\bf
  16}~(2), 290--297.

\bibitem[Hou \& Zhang(2002)]{Hou2002}
{\sc Hou, T.Y.a \& Zhang, P.b} 2002 Convergence of a boundary integral method
  for 3-d water waves. {\em Discrete and Continuous Dynamical Systems - Series
  B\/} {\bf 2}~(1), 1--34, cited By (since 1996)9.

\bibitem[Johannessen \& Swan(2001)]{Johannessen2001}
{\sc Johannessen, T.~B. \& Swan, C.} 2001 A laboratory study of the focusing of
  transient and directionally spread surface water waves. {\em Proceedings:
  Mathematical, Physical and Engineering Sciences\/} {\bf 457}~(2008), pp.
  971--1006.

\bibitem[Kurnia \& van Groesen(2014)]{Kurnia2014}
{\sc Kurnia, R. \& van Groesen, E.} 2014 High order hamiltonian water wave
  models with wave-breaking mechanism. {\em Coastal Engineering\/} {\bf
  93}~(0), 55 -- 70.

\bibitem[Ma(2010)]{Ma2010}
{\sc Ma, Qingwei.} 2010 {\em Advances in numerical simulation of nonlinear
  water waves\/}. Hackensack, NJ: World Scientific.

\bibitem[Ma {\em et~al.\/}(2001)Ma, Wu \& Eatock~Taylor]{Ma2001}
{\sc Ma, Q.~W., Wu, G.~X. \& Eatock~Taylor, R.} 2001 Finite element simulation
  of fully non-linear interaction between vertical cylinders and steep waves.
  part 1: methodology and numerical procedure. {\em International Journal for
  Numerical Methods in Fluids\/} {\bf 36}~(3), 265--285.

\bibitem[Nicholls(1998)]{Nicholls1998}
{\sc Nicholls, David~P.} 1998 Traveling water waves: Spectral continuation
  methods with parallel implementation. {\em Journal of Computational
  Physics\/} {\bf 143}~(1), 224 -- 240.

\bibitem[Park {\em et~al.\/}(2003)Park, Kim, Miyata \& Chun]{Park2003}
{\sc Park, J.~C., Kim, M.~H., Miyata, H. \& Chun, H.~H.} 2003 Fully nonlinear
  numerical wave tank (nwt) simulations and wave run-up prediction around 3-d
  structures. {\em Ocean Engineering\/} {\bf 30}~(15), 1969 -- 1996.

\bibitem[Perlin {\em et~al.\/}(2013)Perlin, Choi \& Tian]{Perlin2013}
{\sc Perlin, Marc, Choi, Wooyoung \& Tian, Zhigang} 2013 Breaking waves in deep
  and intermediate waters. {\em Annual Review of Fluid Mechanics\/} {\bf
  45}~(1), 115--145.

\bibitem[Perlin {\em et~al.\/}(1996)Perlin, He \& Bernal]{Perlin1996}
{\sc Perlin, Marc, He, Jianhui \& Bernal, Luis~P.} 1996 An experimental study
  of deep water plunging breakers. {\em Physics of Fluids\/} {\bf 8}~(9),
  2365--2374.

\bibitem[Phillips(1977)]{Phillips1977}
{\sc Phillips, Owen} 1977 {\em The dynamics of the upper ocean\/}, 2nd edn.
  Cambridge University Press Cambridge ; New York.

\bibitem[Qiao \& Duncan(2001)]{Qiao2001}
{\sc Qiao, Haibing \& Duncan, James~H.} 2001 Gentle spilling breakers: crest
  flow-field evolution. {\em Journal of Fluid Mechanics\/} {\bf 439}, 57-- 85.

\bibitem[Saket {\em et~al.\/}(2017{\natexlab{{\em a\/}}})Saket, Peirson,
  Banner, Barthelemy \& Allis]{Saket2017}
{\sc Saket, Arvin, Peirson, William, Banner, Michael, Barthelemy, Xavier \&
  Allis, Michael} 2017{\natexlab{{\em a\/}}} Wave breaking onset of
  two-dimensional deep-water wave groups in the presence and absence of wind.
  {\em Journal of Fluid Mechanics\/} {\bf 811}, 642--658.

\bibitem[Saket {\em et~al.\/}(2017{\natexlab{{\em b\/}}})Saket, Peirson, Banner
  \& Allis.]{Saket2017a}
{\sc Saket, Arvin, Peirson, William~L., Banner, Michael~L. \& Allis.,
  Michael~J.} 2017{\natexlab{{\em b\/}}} Wave breaking onset of two-dimensional
  wave groups in uniform intermediate depth water. {\em arXiv:1703.04354\/} .

\bibitem[Shemer(2013)]{Shemer2013}
{\sc Shemer, L.} 2013 On kinematics of very steep waves. {\em Natural Hazards
  and Earth System Science\/} {\bf 13}~(8), 2101--2107.

\bibitem[Shemer \& Ee(2015)]{Shemer2015}
{\sc Shemer, L. \& Ee, B.~K.} 2015 Steep unidirectional wave groups -- fully
  nonlinear simulations vs. experiments. {\em Nonlinear Processes in
  Geophysics\/} {\bf 22}~(6), 737--747.

\bibitem[Shemer \& Liberzon(2014)]{Shemer2014}
{\sc Shemer, Lev \& Liberzon, Dan} 2014 Lagrangian kinematics of steep waves up
  to the inception of a spilling breaker. {\em Physics of Fluids\/} {\bf
  26}~(1).

\bibitem[Song \& Banner(2002)]{Song2002}
{\sc Song, Jin-Bao \& Banner, Michael~L.} 2002 On determining the onset and
  strength of breaking for deep water waves. part i: Unforced irrotational wave
  groups. {\em Journal of Physical Oceanography\/} {\bf 32}~(9), 2541--2558.

\bibitem[Stansell \& MacFarlane(2002)]{Stansell2002}
{\sc Stansell, Paul \& MacFarlane, Colin} 2002 Experimental investigation of
  wave breaking criteria based on wave phase speeds. {\em Journal of Physical
  Oceanography\/} {\bf 32}~(5), 1269--1283.

\bibitem[Stokes(1847)]{Stokes1847}
{\sc Stokes, G.~G.} 1847 {On the theory of oscillatory waves}. {\em Trans.
  Camb. Phil. Soc.\/} {\bf 8}, 441--455.

\bibitem[Telles(1987)]{Telles1987}
{\sc Telles, J.C.F} 1987 A self-adaptive co-ordinate transformation for
  efficient numerical evaluation of general boundary element integrals. {\em
  Int. J. Numer. Meth. Engng.\/} {\bf 24}~(5), 959--973.

\bibitem[Tian {\em et~al.\/}(2008)Tian, Perlin \& Choi]{Tian2008}
{\sc Tian, Zhigang, Perlin, Marc \& Choi, Wooyoung} 2008 Evaluation of a
  deep-water wave breaking criterion. {\em Physics of Fluids\/} {\bf 20}~(6),
  066604.

\bibitem[Tulin \& Landrini(2000)]{Tulin2000}
{\sc Tulin, M.P \& Landrini, M.} 2000 Breaking waves in the ocean and around
  ships. In {\em Twenty-Third Symposium on Naval Hydrodynamics\/}, pp.
  713--745. Office of Naval Research, Bassin d'Essais des Carenes, National
  Research Council.

\bibitem[Tulin(2007)]{Tulin2007}
{\sc Tulin, Marshall~P.} 2007 On the transport of energy in water waves. {\em
  Journal of Engineering Mathematics\/} {\bf 58}~(1), 339--350.

\bibitem[West {\em et~al.\/}(1987)West, Brueckner, Janda, Milder \&
  Milton]{West1987}
{\sc West, Bruce~J., Brueckner, Keith~A., Janda, Ralph~S., Milder, D.~Michael
  \& Milton, Robert~L.} 1987 A new numerical method for surface hydrodynamics.
  {\em Journal of Geophysical Research: Oceans\/} {\bf 92}~(C11), 11803--11824.

\bibitem[Xu \& Guyenne(2009)]{Xu2009}
{\sc Xu, Liwei \& Guyenne, Philippe} 2009 Numerical simulation of
  three-dimensional nonlinear water waves. {\em Journal of Computational
  Physics\/} {\bf 228}~(22), 8446 -- 8466.

\bibitem[Xue {\em et~al.\/}(2001)Xue, X\"{u}, Liu \& Yue]{XUE2001}
{\sc Xue, Ming, X\"{u}, Hongbo, Liu, Yuming \& Yue, Dick K.~P.} 2001
  Computations of fully nonlinear three-dimensional wavebody interactions. part
  1. dynamics of steep three-dimensional waves. {\em Journal of Fluid
  Mechanics\/} {\bf 438}, 11--39.

\end{thebibliography}

\end{document}